\documentclass[fleqn,nofootinbib,showkeys,pra]{elsarticle}
\usepackage{setspace}
\newcommand{\R}{{\mathbb{R}}}

\newcommand{\I}{{\mathbb{I}}}
\usepackage{amssymb}
\usepackage{xcolor}

\begin{document}
\title{Scattering  data and bound states of a squeezed double-layer structure
}
\author{Alexander V. Zolotaryuk and Yaroslav Zolotaryuk}
\address
{Bogolyubov Institute for Theoretical Physics, National Academy of
Sciences of Ukraine, Kyiv 03143, Ukraine}


\date{\today}

\begin{abstract}

A  heterostructure composed of two parallel homogeneous layers is studied 
in the  limit as their widths $l_1$ and $l_2$,
 and the distance between them $r$ shrinks to zero
simultaneously. The problem is investigated in one dimension and
the squeezing potential in the Schr\"{o}dinger equation is given by
the  strengths  $V_1$ and $V_2$ 
depending on the layer thickness. A whole class of functions $V_1(l_1)$
and $V_2(l_2)$ is specified by certain limit characteristics
as $l_1$ and $l_2$ tend to zero. The squeezing limit of the scattering 
data $a(k)$ and $b(k)$ derived for the finite system is shown to exist
only if some conditions on the system parameters $V_j$, $l_j$, $j=1,2$, and $r$
take place. These conditions appear as a result of an appropriate cancellation of 
divergences. Two ways of this cancellation are carried out 
and the corresponding two resonance sets in the system parameter space 
are derived. On one of these sets, the existence of non-trivial bound states
is proven in the squeezing limit, including the particular example of the 
squeezed potential in the form of the derivative of Dirac's delta function,
contrary to the widespread opinion on the non-existence of bound
 states in $\delta'$-like systems.  The scenario how a single bound state
 survives in the squeezed system from a finite number of bound states in
 the finite system is described in detail. 

  \end{abstract}

\bigskip

\begin{keyword} 
one-dimensional quantum systems, point interactions,  
 resonant tunneling, bound states. 
\end{keyword}

\maketitle


\section{Introduction}

Exactly solvable models are used in quantum mechanics to describe the properties
of realistic physical systems such as scattering coefficients, bound states, etc.
in closed form using elementary functions. A whole class of these models is 
represented by Schr\"{o}dinger operators with singular zero-range potentials
defined on the sets consisting of isolated points. In the literature these 
operators are referred to as 
contact or {\it point interaction} models (see books \cite{do1,do2,a-h,ak} for
details and references).

Beginning from the pioneering paper of Berezin and Faddeev \cite{bf}, 
who suggested to treat formal Schr\"{o}dinger operators with singular perturbations
as mathematically well-defined objects using the theory of self-adjoint extensions of 
symmetric operators, a whole family of point interactions have been constructed
in  numerous works
(see, e.g., publications \cite{a-h,ak,k,adk,an,n1,n2,acf,ce,gnn,bn,kp,n-r}, 
a few to mention). Thus, an interesting approach based on the integral 
form of the one-dimensional Schr\"{o}dinger equation
has been developed by Lange \cite{l1,l2} with some revision of 
 Kurasov's theory \cite{k}. 

Another approach is
to approximate the singular potential part of the formal Schr\"{o}dinger equation
by regular finite-range functions and to study the convergence in 
the norm-resolvent topology \cite{s,enz}. Within this procedure, 
 the point interactions such as a $\delta'$-interaction or a $\delta'$-potential
 (the difference between these interactions is explained in \cite{bn}),
 which are more singular than a $\delta$-potential, are realized in the
 squeezing limit. Thus, Cheon and Shigehara \cite{cs} were
the first who developed an approach how to construct a whole family of point
interactions from shrinking three separated Dirac's delta potentials to one point,
using various potential strengths.  Later, Exner, Neidhardt and Zagrebnov
\cite{enz} have rigorously shown that  there exists 
 a family Schr\"{o}dinger operators with regular potentials,
which approximates the $\delta'$-interaction Hamiltonian in the norm-resolvent sense,
and this approximation  has non-trivial convergence properties.
As concerns the $\delta'$-potential,  using 
a general regularization of the derivative of Dirac's delta function
$\beta \delta'(x)$ ($\beta \in \R$ is a strength constant)
as a potential part in the one-dimensional Schr\"{o}dinger equation, \v{S}eba \cite{s} 
has constructed the point interaction which is opaque {\it everywhere}
on the $\gamma$-axis. However, for some particular regularizing sequences of $\delta'(x)$,
a perfect reflection was observed \cite{c-g,zci,tn,z10pla,zz14} 
{\it almost everywhere} on the $\beta$-axis. 
In other words, a partial transmission of particles through the potential
$\beta \delta'(x)$ was shown to occur at certain isolated points $\{ \beta_n \}$ 
forming a set of Lebesgue's measure zero depending on the regularizing sequence.  
In general terms, for a whole class of regularizing sequences 
of the potential $\beta \delta'(x)$,
Golovaty and coworkers \cite{gm,gh,gh1,g2} have rigorously proved the 
existence of a {\it resonance set} in the $\beta$-space and its
dependence on the regularization. On this set, they constructed  the algorithm 
for computing the matrix that connects  the two-sided boundary conditions
at the point of singularity and allows to calculate 
the reflection-transmission coefficients. 

 Besides `pure' single-point interactions,  more specified models, 
which are important in applications, have been elaborated. Thus, 
the potential  $V(x) = \alpha \delta(x) + \beta \delta'(x)$, $\alpha <0$, 
$\beta \in \R$, has been used by Gadella {\it et al.}
as a perturbation of the background 
potentials such as the harmonic oscillator \cite{ggn} or
the infinite square well \cite{gggm}. 
The spectrum of  a one-dimensional V-shaped quantum well perturbed by three types of 
 a point impurity as well as 
three solvable two-dimensional systems (the isotropic harmonic oscillator,
a square pyramidal potential and their combination) perturbed 
by a point potential centered at the
origin has  been studied by Fassari and coworkers in the recent papers 
\cite{fggn1,fggn2,f-r,a-f-r}. Some aspects of the $\delta'$-interaction, 
   its approximations by local and non-local potentials as well as
   its combination with background potentials, have been investigated in
the series of works \cite{enz,an00,an07,an13,afr1,afr2}.

Two-point interactions are also important in applications. 
The resonant tunneling through double-barrier scatters is still 
an active area of research for the applications to
nanotechnology \cite{knt1,knt2}.   
Tunneling times in double-barrier point potentials  
 and the associated questions such as the generalized version of the Hartman effect 
have been studied in  \cite{clm,lmn,clmn,llmn1,llmn2}.  Another important 
aspect regarding the application of double-point potentials is the Casimir 
effect that arises in the behavior
of the vacuum energy between two homogeneous parallel plates. 
For the interpretation of this effect,
 Mu\~{n}oz-Casta\~{n}eda and coworkers \cite{ag-am-c,am-c1,gm-c,am-c2,m-cgm} 
 reformulated the theory of self-adjoint extensions of symmetric operators over
bounded domains in the framework of quantum field theory. Particularly, they 
have calculated the vacuum energy and identified which boundary conditions 
generate attractive or repulsive Casimir forces between the plates.

The present paper is devoted to the investigation of a planar 
heterostructure composed of two
 extremely thin layers separated by a small distance in the limit as
both the layer thickness and the distance between the layers  simultaneously
shrink to zero.  The electron motion in this system is supposed to be 
 confined in the longitudinal
direction (say, along the $x$-axis) being perpendicular to the transverse planes 
where electronic motion is free. 
The three-dimensional Schr\"{o}dinger equation of such a structure can be separated
into longitudinal and transverse parts, resulting in  the reduced 
stationary one-dimensional Schr\"{o}dinger equation 
\begin{equation}
-\, \psi''(x) +V(x)\psi(x)= E\psi(x)
\label{1}
\end{equation}
with respect to the longitudinal component of the
wave function $\psi(x)$ and the electron energy $E$. Here, $V(x)$ is 
a potential for electrons to be specified below 
for a double-layer structure and the prime stands for
the differentiation over $x$. 
Concerning numerical computations, we note that equation (\ref{1}) 
has been written in the system for which $\hbar^2/2m^* =1$ where
$m^*$ is an effective electron mas. In the calculations we set
$m^* =0.1 \, m_e$ where $m_e$ is the free electron mass, 
so in our case 1 eV = 2.62464 nm$^{-2}$.

The potential $V(x)$ in equation (\ref{1}) 
is understood as  a sequence of regular potentials that shrinks to a point.
It is not necessary for this sequence to have any well-defined 
limit even in the sense of distributions   
for producing a point interaction. Thus, ignoring any condition for the 
existence of a distributional limit, one can extensively enlarge 
the family of regularizing sequences. Thus, for a double-layer structure,
the resonance set consist of curves that covers the corresponding resonance set
for the $\delta'$-potential formed by the points lying on these curves
 \cite{z17,z18aop,z19}. Another aspect of this approach is the 
 possibility to investigate in a squeezing limit 
 the resonant-tunneling transmission through biased potentials
\cite{zz15pla,zz15jpa,ztz}.

The goal of this article is to construct the family of point interactions 
from a double-layer potential as this potential shrinks to a point.
The approach is based on the  analysis of the
scattering data $a(k)$ and $b(k)$ \cite{d-m}, ($k:= \sqrt{E}\,$),
in a squeezing limit. In general, the functions $a(k)$ and $b(k)$
diverge in this limit, however, at some constraints on the system 
parameters, a cancellation of divergences may happen, leading
to the existence of finite values for the scattering data.

The definition of the scattering data $a(k)$ and $b(k)$ is given 
in the next section. Here, the relationship of these data with
the transmission matrix for a single-layer structure is 
discussed. The scattering data for a double-layer structure 
are present in section~3. Two types of resonance sets are found
 in the next section 
as the result of cancellation of divergences. The behavior of 
bound states in a squeezing limit is described in section~5.
In the next section, the three-scale power-connecting parametrization
of the double-layer potential is applied for the numerical illustration
of the behavior of bound states and wave functions under the squeezing procedure.
Finally, some concluding remarks are discussed in section~7.

\section{Preliminaries}

In this section we present the definitions of 
the scattering functions $a(k)$ and $b(k)$, and some relations
regarding the transmission matrix, reflection-transmission coefficients
and bound states.

\subsection{Monodromy matrix}

For equation (\ref{1})
we define two sets of linearly independent solutions to this equation
as follows
\begin{equation}
\begin{array}{ll} \phi_1(x) \sim {\rm e}^{-\,{\rm i}kx},~~~
\phi_2(x) \sim {\rm e}^{{\rm i}kx}~~& \mbox{as}~~x \to -\,\infty ,\\
\psi_1(x) \sim {\rm e}^{-\,{\rm i}kx},~~~
\psi_2(x) \sim {\rm e}^{{\rm i}kx}~~& \mbox{as}~~x \to + \,\infty. \end{array}
\label{2}
\end{equation}  
This pair of solutions can be coupled through the monodromy matrix
${\bf M}$ as 
\begin{equation}
\mbox{col}\{ \phi_1(x),\, \phi_2(x)\} = {\bf M} \,
\mbox{col}\{ \psi_1(x),\, \psi_2(x)\},
\label{3}
\end{equation}
where 
\begin{equation}
{\bf M} = \left( \begin{array}{lr} a(k) ~~~b(k) \\
b^*(k)~~a^*(k) \end{array} \right),~~~\det{\bf M} = |a|^2 - |b|^2 =1.
\label{4}
\end{equation}
On the other hand, one can write
\begin{equation}
\mbox{col}\{ \psi_1(x),\, \psi_2(x)\} = {\bf M}^{-1} 
\mbox{col}\{ \phi_1(x),\, \phi_2(x)\},
\label{5}
\end{equation}
where 
\begin{equation}
{\bf M}^{-1} = \left( \begin{array}{lr} ~~a^*(k) ~~~- b(k) \\
-\, b^*(k)~~~~~~a(k) \end{array} \right).
\label{6}
\end{equation}

\subsection{Reflection-transmission coefficients and  equations for bound states}

The reflection-transmission coefficients for an incident plane 
wave from the right are defined by
\begin{equation}
\psi(x) = \left\{ \begin{array}{ll}
 T_r {\rm e}^{-\,{\rm i} kx}  &  \mbox{as}~~ x \to -\,\infty ,  \\
  {\rm e}^{-{\rm i} kx} + R_r\, {\rm e}^{{\rm i} kx} & \mbox{as}
  ~~ x \to +\, \infty  \end{array} \right. 
\label{7}
\end{equation}
and, from the comparison with the 
expression of the matrix ${\bf M}$ given by equations (\ref{3}) and
(\ref{4}), we obtain
\begin{equation}
R_r = b(k)/a(k),~~~T_r = 1/a(k) .
\label{8}
\end{equation}
Then the equation for bound states reads $a( {\rm i}\kappa) =0$. 

Similarly,  the reflection-transmission coefficients for an incident plane 
wave from the left are defined by
\begin{equation}
\psi(x) = \left\{ \begin{array}{ll}
 {\rm e}^{{\rm i} kx} + R_l\, {\rm e}^{-{\rm i} kx} &  \mbox{as}~~
x \to -\,\infty ,  \\
 T_l\, {\rm e}^{{\rm i} kx} & \mbox{as}~~x \to   +\, \infty 
  \end{array} \right. 
\label{9}
\end{equation}
and, from the comparison with the 
expression for the matrix ${\bf M}^{-1}$ given by equations (\ref{5})
and (\ref{6}), we obtain
\begin{equation}
R_l =-\, b^*/a,~~~T_l = 1/a .
\label{10}
\end{equation}
In this case the equation for bound states reads $a^*( -\, {\rm i}\kappa) =0$. 

\subsection{Transmission matrix and its relation to a monodromy matrix}

 For a finite-range system supported on the interval $x_1 \le x \le x_2$,
for which $V(x) \equiv 0$ on the intervals 
$-\,\infty < x <x_1$ and $x_2 <x < +\, \infty$, the 
 transmission matrix $\Lambda$ is defined by the matrix equation
\begin{equation}
 \left( \begin{array}{lr} \psi(x_2) \\
\psi'(x_2) \end{array} \right) =\Lambda \left( \begin{array}{lr} \psi(x_1) \\
\psi'(x_1) \end{array} \right),~\Lambda = \left( \begin{array}{lr} 
\lambda_{11} ~~~\lambda_{12} \\ \lambda_{21}~~~\lambda_{22} \end{array} \right)
, ~\det\Lambda =1,
\label{11}
\end{equation}
for both positive- ($k >0$) and negative-energy ($k ={\rm i} \kappa $) solutions
of equation (\ref{1}).
Note that in both these cases, the elements 
$\lambda_{ij}$'s are real-valued functions of $k$ or ${\rm i}\kappa$.  

In general, the $\Lambda$-matrix can be defined as follows.
Let $u(x)$ and $v(x)$ be linearly independent solutions of the
Schr\"{o}dinger equation (\ref{1}) for a layer placed on the interval 
$x_1 \le x \le x_2$. Then one can express the elements $\lambda_{ij}$ in terms  
of the initial values of these solutions in the following form: 
\begin{equation}
\begin{array}{llll}
\lambda_{11}= W(x_1)^{-1}  \left[ u(x_2)v'(x_1) - u'(x_1)v(x_2)\right],\\
\lambda_{12}= W(x_1)^{-1} \left[ u(x_1)v(x_2) - u(x_2)v(x_1)\right],\\
\lambda_{21}= W(x_1)^{-1} \left[ u'(x_2)v'(x_1) - u'(x_1)v'(x_2)\right],\\
\lambda_{22}= W(x_1)^{-1} \left[ u(x_1)v'(x_2) - u'(x_2)v(x_1)\right],
\end{array} \label{12}
\end{equation}
where $W(x_1) =  u(x_1)v'(x_1) - u'(x_1)v(x_1) =W(x)$, $x_1 \le x \le x_2$,
 is the Wronskian of equation (\ref{1}). Equations (\ref{12}) are simplified if the
 solutions $u(x)$ and $v(x)$ satisfy the initial conditions at one of the
 ends $x=x_1$ or $x=x_2\,.$ Let 
 \begin{equation}
 u(x_1) =1,~~u'(x_1)=0,~~v(x_1)=0,~ ~v'(x_1)=1.
 \label{13}
 \end{equation}
Then, using these values in equations (\ref{12}), one immediately finds
\begin{equation}
\Lambda = \left( \begin{array}{lr} u(x_2)~~~v(x_2) \\
u'(x_2)~~ v'(x_2) \end{array} \right), ~~~\det \Lambda =1.
\label{14}
\end{equation}

Similarly, the solution to equation (\ref{1}) 
can also be expressed in terms of the functions $u(x)$ and $v(x)$
defined by initial conditions (\ref{13}). Let $\psi(x_1)$ and $\psi'(x_1)$
be boundary conditions at $x=x_1$. Then this solution on the interval 
$x_1 \le x \le x_2 < \infty$  reads
\begin{equation}
\psi(x)= \psi(x_1)u(x) + \psi'(x_1)v(x).
\label{15}
\end{equation}

The scattering data $a(k)$ and $b(k)$ can be  expressed via the elements of
the $\Lambda$-matrix defined on the interval $x_1 \le x \le x_2\,$. 
To this end, consider representation (\ref{2})
for the solution $\phi_1(x)$, writing  
\begin{equation}
\phi_1(x) = \left\{ \begin{array}{ll}
 {\rm e}^{-{\rm i} kx},   & ~~  - \infty < x <x_1 , \\
 a(k){\rm e}^{-\,{\rm i} kx} + b(k){\rm e}^{{\rm i} kx}, &  ~~~ x_2 < x < +\,\infty  . 
 \end{array} \right. 
\label{16}
\end{equation}
Inserting the values of this function and its derivatives at $x =x_1$ and $x=x_2$
into the equations
\begin{equation}
\begin{array}{ll}
\phi_1(x_2)= \lambda_{11}\phi_1(x_1) + \lambda_{12}\phi_1'(x_1),\\
\phi'_1(x_2)= \lambda_{21}\phi_1(x_1) + \lambda_{22}\phi_1'(x_1),\end{array}
\label{17}
\end{equation}
and solving the resulting pair of equations with respect to $a(k)$ and $b(k)$,
we immediately find 
\begin{equation}
a(k) = \frac{D(k)}{2}  {\rm e}^{{\rm i} k (x_2 - x_1)},~~~
b(k)=\frac{p(k)- {\rm i}q(k)}2 {\rm e}^{-\, {\rm i} k(x_1 + x_2)},
\label{18}
\end{equation}
where
\begin{equation}
\!\!\!\!
D :=  \lambda_{11} + \lambda_{22} -\, {\rm i} (k \lambda_{12} -
k^{-1}\lambda_{21}),~~p := \lambda_{11} - \lambda_{22}\,,~~
q := k\lambda_{12} + k^{-1}\lambda_{21}. ~~~
\label{19}
\end{equation}
Using the equation $\det\Lambda  =1$, one can derive the  equality 
$|D|^2 =4 +p^2 +q^2$ and, as a result,  for positive-energy solutions 
($k =\sqrt{E} $), the relation  $\det{\bf M}=1$  holds true. 
In a squeezed limit, one can set $x_1 \to -\,0$ and 
$x_2 \to +\, 0$, so that $a(k) =D/2$ and $b(k)=(p-{\rm i}q)/2$. 

The equation for bound states reads $a({\rm i}\kappa)=0$ and, as a result, 
from (\ref{18}) with  (\ref{19}) we get
 the following general equation given in terms of
the $\Lambda$-matrix elements: 
\begin{equation}
 \lambda_{11}(\kappa) + \lambda_{22}(\kappa)
+ \kappa \lambda_{12}(\kappa) + \kappa^{-1}\lambda_{21}(\kappa) =0.
\label{20}
\end{equation}
This is a general equation given in terms of the elements of the $\Lambda$-matrix.

\section{Transmission matrix, 
scattering data and wave functions for a double-layer potential}

Consider the system consisting of two separated layers described by the 
piecewise constant potential, which is defined on the whole axis as follows
 \begin{equation}
 V(x)= \left\{ \begin{array}{lll}  V_1, & 0 < x < l_1\,, \\
 V_2,   & l_1 +r <x<  l_1+l_2+r, \\
 ~0\,, & -\,\infty < x < 0 ,~ l_1 < x < l_1 +r, 
 ~l_1 + l_2 +r < x< \infty . \end{array} \right.
 \label{21}
 \end{equation}
Here $V_j \in \R$  (barrier if $V_j > 0$ or well if $V_j < 0$),  $l_j >0$ 
(layer thickness) and $r >0$ (distance between layers), $j=1,2$. 

 On each of the three intervals $0 < x< l_1$, $l_1 < x < l_1 +r$ and 
$l_1 + r < x < l_1 +r +l_2$,  the corresponding transmission matrices, 
denoted by $\Lambda_1$, $\Lambda_0$, $\Lambda_2$, respectively, 
can be written immediately on the basis of general formula (\ref{14}). 
Thus, setting $x_1 =0$ and $x_2 = l_1$ for $\Lambda_1$, 
$x_1 = l_1$ and $x_2 =l_1 + r $ for  $\Lambda_0$,
  $x_1 = l_1 +r$ and $x_2 =l_1 + r +l_2$ for $\Lambda_2$, we have 
  \begin{equation}
  \!\!\!\!\!\!\!\!\!\!\!\!\!
\Lambda_j = \left( \begin{array}{lr} 
\cos(k_jl_j)  ~~~ ~~k_j^{-1}\sin(k_jl_j) \\  - k_j \sin(k_jl_j)~~~~
\cos(k_jl_j) \end{array} \right),~\Lambda_0 = \left( \begin{array}{lr} 
\cos(k r)  ~~~k^{-1}\sin(k r) \\ - k\sin(k r) 
~~~~~\cos(k r) \end{array} \right),
\label{22}
\end{equation} 
 with $ k_j = \sqrt{k^2 - V_j}\,,$ $j=1,2$.  
Then the total transmission matrix is the product 
$\Lambda =\Lambda_2 \Lambda_0 \Lambda_1 $ with the elements
\begin{equation} 
\!\!\!\!\!\!\!\!
\begin{array}{llll}
\lambda_{11}  = \left[ \cos(k_1l_1) \cos(k_2l_2) 
-(k_1/k_2)\sin(k_1l_1) \sin(k_2l_2) \right] \cos(kr) \\
~~~~\, - \left[ (k_1/k) \sin(k_1l_1) \cos(k_2l_2) +   
(k/k_2) \cos(k_1l_1) \sin(k_2l_2) \right] \sin(kr), \\
\lambda_{12}  = \left[(1/k_1) \sin(k_1l_1) \cos(k_2l_2)
+ (1/k_2)\cos(k_1l_1) \sin(k_2l_2) \right] \cos(kr) \\
 ~~~~\, + \left[ (1/k) \cos(k_1l_1) \cos(k_2l_2)
  -  (k/k_1k_2) \sin(k_1l_1) \sin(k_2l_2) \right] \sin(kr), \\
\lambda_{21}  = - \left[ k_1 \sin(k_1l_1)\cos(k_2l_2)
+ k_2 \cos(k_1l_1)\sin(k_2l_2) \right] \cos(kr) \\
~~~~\, - \left[ k \cos(k_1l_1) \cos(k_2l_2)  
-  (k_1k_2/k) \sin(k_1l_1) \sin(k_2l_2) \right] \sin(kr), \\
\lambda_{22}  =  \left[ \cos(k_1l_1) \cos(k_2l_2)
-(k_2/k_1)\sin(k_1l_1) \sin(k_2l_2) \right] \cos(kr) \\
~~~~ \,- \left[ (k/k_1) \sin(k_1l_1)\cos(k_2l_2)
 +   (k_2/k) \cos(k_1l_1) \sin(k_2l_2) \right] \sin(kr). \end{array}
\label{23}
\end{equation}

Setting $x_1 =0$ and $x_2= l_1 +l_2 +r$ in equations (\ref{18}) and (\ref{19}),
where the $\Lambda$-matrix elements are given by equations (\ref{23}),
 we find the scattering data:
\begin{eqnarray}
\!\!\!\!\!\!\!\!\!\!\!\!\!\!\!\!\!\!\!\!\!\!\!\!
&& { a(k) \over \cos(k_1l_1)\cos(k_2l_2)} =
\left\{ {\rm e}^{- {\rm i}kr} - {{\rm i} \over 2}\! \left[ 
\left({k \over k_1}+ {k_1 \over k}\right)t_1 +
\left({k \over k_2}+  {k_2 \over k}\right)t_2\right] {\rm e}^{-{\rm i}kr}\right.
 \nonumber \\
\!\!\!\!\!\!\!\!\!\!\!\!\!\!\!\!\!\!\!\!\!\!\!\!
&& + \left. {1 \over 2} \left[ {\rm i}\! \left({k^2 \over k_1k_2}
 + {k_1k_2 \over k^2}\right)  \sin(kr)  
 - \left({k_1 \over k_2}+  {k_2 \over k_1}\right) \cos(kr)\right]t_1t_2
  \right\}  {\rm e}^{{\rm i}k(l_1 +l_2+r)}, 
\label{24}
\end{eqnarray}
\begin{eqnarray}
\!\!\!\!\!\!\!\!\!\!\!\!\!\!\!\!\!\!\!\!\!\!\!
&& { b(k) \over \cos(k_1l_1)\cos(k_2l_2)} =
  {{\rm i} \over 2} \left\{ \left( {k_1 \over k} 
- {k \over k_1}\right)t_1\,{\rm e}^{{\rm i}kr} +  \left( {k_2 \over k} 
- {k \over k_2}\right)t_2\,{\rm e}^{- {\rm i}kr} \right.\nonumber \\
\!\!\!\!\!\!\!\!\!\!\!\!\!\!\!\!\!\!\!\!\!\!\!
&& +\left.  \left[ \left( {k^2 \over k_1k_2} - {k_1k_2 \over k^2}\right) \sin(kr) 
  + {\rm i}\left({k_1 \over k_2}   -{k_2 \over k_1}\right)\cos(kr)\right] 
 t_1t_2 \right\}  {\rm e}^{-{\rm i}k(l_1 +l_2 +r)},
 \label{25}
\end{eqnarray}
where $t_j := \tan(k_jl_j),$ $j=1,2.$

A family of {\it one}-point interactions can be realized from 
equation (\ref{1}) with potential (\ref{21}) if all the three size parameters 
$l_1, l_2$ and $r$ converge to the origin $x=0$, 
whereas the values $|V_1|$ and $|V_2|$ (and therefore $|k_1|$ and $|k_2|$) 
must tend  to infinity. 
However,  the arguments of trigonometric functions 
$k_1l_1$ and $k_2l_2$ in equations (\ref{23}) must be finite 
(including  zero) in the limit
as $l_1, l_2 \to 0$. Therefore one can consider a whole family of functions
$V_1 = V_1(l_1)$ and $V_2=V_2(l_2)$, for which the arguments will be finite.
To this end,  let us define the function set ${\cal G} = {\cal G}_1\times{\cal G}_2$
with
\begin{equation}
{\cal G}_j := \{ V_j(l_j) \mid \lim_{l_j \to 0 }|V_j(l_j)|^{1/2}\,l_j = c_j \},~~~j=1,2,
\label{26}
\end{equation}
where  the constants $c_1$ and $c_2$ are finite or zero ($0 \le c_j < \infty$).
Since the constants $c_j$'s can be either non-zero or zero, the four  cases of
 ${\cal G}$ should 
be considered separately:  ${\cal G}_{11}(c_1 >0,\, c_2 >0)$,  
${\cal G}_{01}(c_1 =0, \, c_2 >0)$, ${\cal G}_{10}(c_1 >0,\, c_2 =0)$,
${\cal G}_{00}(c_1 =0, \,c_2 =0)$. 

Let us analyze first the asymptotic behaviour of the elements $\lambda_{11}$ and 
$\lambda_{22}$ in  (\ref{23}) as  $l_1, l_2, r \to 0$. They will 
 be finite and non-zero  if 
\begin{equation}
\!\!\!\!\!\!\!\!\!\!\!\!\!\!\!\!
\begin{array}{llll}
|V_1(l_1)/V_2(l_2)| \to \mbox{const.}>0 & \mbox{for}~{\cal G}_{11}, \\
|V_1(l_1)|\,|V_2(l_2)|^{-1/2} \, l_1 \to \mbox{const.} \ge 0,~~
|V_2(l_2)|^{1/2}\,l_1  \to \mbox{const.} \ge 0 & \mbox{for}~{\cal G}_{01}, \\
|V_1(l_1)|^{- 1/2}\,|V_2(l_2)|\,l_2  \to \mbox{const.} \ge 0,~~
|V_1(l_1)|^{1/2}l_2  \to \mbox{const.} \ge 0 & \mbox{for}~{\cal G}_{10}, \\
|V_1(l_1)|\,l_1l_2  \to \mbox{const.} \ge 0 ,~~|V_2(l_2)|\,l_1l_2  \to
\mbox{const.} \ge 0  & \mbox{for}~{\cal G}_{00},
\end{array} 
\label{27}
\end{equation}
and  the distance $r$ shrinks to zero
 sufficiently fast compared with  the squeezing of $l_1$ and $l_2$,
so that each of the following  expressions:
\begin{equation}
\!\!\!\!\!\!\!\!\!\!
\begin{array}{llll}
|V_1(l_1)|^{1/2}\, r, ~|V_2(l_2)|^{1/2}\,r~~({\cal G}_{11}),~~&
 |V_1(l_1)|\,l_1 r, ~|V_2(l_2)|^{1/2}\,r ~~({\cal G}_{01}), \\ 
 |V_1(l_1)|^{1/2} \,r, ~ |V_2(l_2)|\,l_2 r~~~\,({\cal G}_{10}),~~&
 |V_1(l_1)|\,l_1 r, ~ |V_2(l_2)|\,l_2 r~~~\,({\cal G}_{00}),
\end{array}~~
\label{28}
\end{equation}
must converge  to an arbitrary constant or zero. 
Using the definition of the ${\cal G}$-sets,
from conditions (\ref{27}) one can 
derive asymptotic relations between $l_1 \to 0$ and $l_2 \to 0$
in the form of ratios
\begin{equation}
\!\!\!\!\!\!\!\!\!\!\!\!\!\!
{l_1 \over l_2}~({\cal G}_{11}), ~ { |V_1(l_1)|\,l_1 \over |V_2(l_2)|^{1/2}}
~({\cal G}_{01}),~ { |V_1(l_2)|\,l_2 \over |V_1(l_1)|^{1/2}}   ~({\cal G}_{10}), ~
{ |V_1(l_1)|l_1 \over  |V_2(l_2)|l_2}~({\cal G}_{00}) \to \mbox{const.} >0,~
\label{29}
\end{equation}
which are required  to 
converge to arbitrary {\it non-zero} constants. Under these conditions,
 one can check that limits (\ref{27}) indeed take place.
 For instance,  for the ${\cal G}_{01}$-set, we have  $|V_2|^{1/2}l_1 =
  |V_1|\, l_1^2 \,\, |V_2|^{1/2}/ |V_1|\,l_1 \to 0 $ as required. Next, e.g.,
  owing to  the last ratio in (\ref{29}), $|V_1|\,l_1l_2 =\mbox{const.}|V_2|\,l_2^2 \to 0$
  for the ${\cal G}_{00}$-set.    Having the asymptotic relative behavior of $l_1$ 
and $l_2$ given by expressions (\ref{29}),   now one can derive  asymptotic relations
 $r=r(l_1)$ or $r=r(l_2)$ fulfilling limits (\ref{28}). Thus, using 
 these limits as well as the definition of the ${\cal G}$-sets,
 under the requirement that  expressions
\begin{equation}
{r \over l_1} ~({\cal G}_{11}),~~~ 
{r \over l_2}~({\cal G}_{01}) ,
 ~~ ~{r \over l_1}~({\cal G}_{10}), ~~~|V_1(l_1)|\,l_1 r ~({\cal G}_{00})
 \to \mbox{const.} \ge 0,
 \label{30}
\end{equation}
have to converge to arbitrary (positive) constants  or  zero, 
we are convinced  that conditions (\ref{28}) hold true.

Thus, under the conditions imposed on the asymptotic 
behavior of the parameters $l_1, l_2, r$ as they 
converge to the origin in the way described by  relations 
(\ref{29}) and (\ref{30}) having finite limits, the limit  
$\Lambda$-matrix elements $\lambda_{11}$ and $\lambda_{22}$ are 
finite and non-zero. For each of the 
${\cal G} = \{ {\cal G}_{11},\, {\cal G}_{01},\,{\cal G}_{10},\,{\cal G}_{00} 
\}$-sets, all these convergence ways  $\{ l_1, l_2, r\} =: \gamma \to 0$ 
 can be interpreted respectively as  {\it pencils of paths}
 $\Gamma =\{ \Gamma_{11}, \, \Gamma_{01}, \, \Gamma_{10}, \, \Gamma_{00} \}$ 
 in the $\{l_1, l_2, r\}$-space with the vertex located 
at the origin, so that any path $\gamma \in \Gamma$ provides   finite non-zero 
limits of $\lambda_{11}$ and $\lambda_{22}$.
Since $|k_j| \to \infty $ in the limit as $\gamma \to 0$, 
 we have $\lambda_{12} \to 0$, while
 $\lambda_{21}$ in general diverges as $\gamma = \{ l_1, l_2, r\} \to 0$. Therefore
 the  two-sided boundary conditions on the wave function 
$\psi(x)$ in this limit are of the Dirichlet type:
\begin{equation}
\psi(\pm 0)=0~~~\mbox{if}~~|\lambda_{21}| \to \infty.
\label{31}
\end{equation}
This is a particular case of the second part of the  Albeverio-D\c{a}browski-Kurasov
(ADK) theorem \cite{adk}, regarding {\it separated}  point interactions.
However, under some constraints on the limit (resonance) constants of the expressions
listed in (\ref{29}) and (\ref{30}), the squeezing limit of 
$\lambda_{21}$ may be finite or zero and in this case we are dealing with
the first part of the ADK theorem, when the resulting point interactions
are {\it non-separated}. The corresponding jump conditions at $x=0$
on the wave function $\psi(x)$ and its derivative $\psi'(x)$ will be given below
in an explicit form.

Finally, on the basis of equation (\ref{15}),  the wave function as a solution 
of equation (\ref{1}) with potential (\ref{21}) 
 can easily be computed explicitly using the linearly independent solutions
 obeying initial conditions (\ref{13}) on each of the intervals:
 $0 \le x \le l_1$, $l_1 \le x \le l_1 +r$, $l_1 +r \le x \le l_1 +l_2 +r$.
  Thus, on the whole axis $-\, \infty < x <
 \infty$, the positive-energy solution describing an incident plane wave from the right
 with asymptotic representation (\ref{16}) reads
\begin{equation}
\!\!\!\!\!\!\!\!\!\!\!\!\!\!
\phi_1(x) = \left\{\begin{array}{lllll}
\varphi_l(x): = {\rm e}^{- {\rm i}kx}, & -\, \infty < x <0,\\
\varphi_1(x):= \cos(k_1x) - {\rm i}(k/k_1)\sin(k_1x), & 0< x < l_1,\\
\varphi_0(x):= \varphi_1(l_1)\cos[k(x-l_1)] \\
    ~~~~~~ +k^{-1}\varphi'_1(l_1)\sin[k(x-l_1)],
& l_1 < x < l_1 +r, \\
\varphi_2(x):= \varphi_0(l_1+r)\cos[k_2(x-l_1-r)] & \\
~~~  + k_2^{-1}\varphi'_0(l_1+r)\sin[k_2(x-l_1-r)],
& l_1+r < x < l_1+l_2  +r,\\
\varphi_r(x):= a(k){\rm e}^{- {\rm i}kx} + b(k){\rm e}^{{\rm i}kx}, &
l_1+l_2 +r <x <  \infty.
\end{array}\right. 
\label{32}
\end{equation}
Setting in these formulas $k = {\rm i}\kappa$, we obtain the wave function
corresponding to the discrete spectrum and the bound energy levels $\kappa$'s
are found from the equation $a({\rm i}\kappa)=0$. In this case, the last
formula in (\ref{32}) reads $\varphi_r(x)=  b({\rm i}\kappa)
\exp(-\kappa x)$.

\section{Resonance sets for the existence of scattering data 
and a bound state in a squeezed limit}

Consider an asymptotic representation of equations (\ref{24}) and (\ref{25})
in the limit  as $\gamma \to 0$. 
 Expanding $\exp(\pm {\rm i}kr) = 1 \pm {\rm i}kr + {\cal O}(r^2)$ 
in the curly brackets of these equations and omitting  the terms which 
vanish in  the limit as $\gamma \to 0$, one can write  
the following asymptotic representation of the scattering data:
\begin{equation} 
 a(k)  \simeq   \frac12 \left[ z_1(k) + z_2(k)  - {{\rm i}  \over k} \Delta (k) 
  \right]\! \cos(k_1l_1)\cos(k_2l_2) \, {\rm e}^{{\rm i}k(l_1 + l_2 +r)},  
\label{33} 
\end{equation}
\begin{equation}
 b(k)  \simeq  \frac12 \left[ z_1(k) - z_2(k)  +{{\rm i}\over k}\Delta(k) \right] 
\! \cos(k_1l_1)\cos(k_2l_2)\,{\rm e}^{-{\rm i}k(l_1 + l_2 +r)},
\label{34}
\end{equation}
where
\begin{equation}
z_1(k) := 1 -k_1t_1r -(  k_1 / k_2)t_1t_2 \,,~~z_2(k) := 1 -k_2t_2r 
-(  k_2 / k_1)t_1t_2 
\label{35}
\end{equation}
and
\begin{equation}
 \Delta(k) := k_1t_1 + k_2t_2   -\, k_1t_1k_2t_2 r.
\label{36}
\end{equation}
The function  $\Delta(k)$ is the most singular term and it diverges in general 
 as $\gamma \to 0$. However, under certain conditions imposed on the system
 parameters, this term may be finite if a squeezing procedure is carried out 
 in a proper way. This happens if a  cancellation of divergences occurs
 in this term. Indeed, such a cancellation procedure does exist
 if a squeezed limit is arranged in the two ways as follows. The first way is to  
set $\Delta(k) =0$, where the distance $r$ participates in the cancellation.
The second way is to suppose that  $r \to 0$ sufficiently fast, available to
suppress the divergent product $k_1k_2$ in the last term of (\ref{36}).
In this case, only the first two terms in (\ref{36}) must be canceled out.

\subsection{The first way of the cancellation of divergences}

Using the condition $\Delta(k) =0$ and the expressions for $z_1(k)$ and 
$z_2(k)$ given by equations (\ref{35}), one can establish that
\begin{eqnarray}
&& \left(1 -k_1t_1r - {k_1 \over k_2}t_1t_2 \right)\!\cos(k_1l_1) \cos(k_2l_2)
= -\,{k_1 \sin(k_1l_1) \over k_2 \sin(k_2l_2) } \nonumber \\
&& ={\cos(k_1l_1) - k_1r \sin(k_1l_1) \over  \cos(k_2l_2)} =
{\cos(k_1l_1) \over \cos(k_2l_2) - k_2r \sin(k_2l_2) }= : \theta(k)
\label{37}
\end{eqnarray}
and furthermore
\begin{equation}
\left(1 -k_2t_2r - {k_2 \over k_1}t_1t_2 \right)\!\cos(k_1l_1)\cos(k_2l_2)
=  \theta^{-1}(k).
\label{38}
\end{equation}
In order to prove equations (\ref{37}), we rewrite the condition 
$\Delta(k) =0$ [see equation  (\ref{36})] as the following three equations: 
 \begin{equation}
k_1t_1r = 1 + {k_1t_1 \over k_2t_2}\,, ~~{k_1 \over k_2 } = k_1t_2r -{t_2 
\over t_1}\,,~~k_1 = {k_2t_2 \over t_1 (k_2t_2r -1)}\,.
\label{38a}
 \end{equation}
Inserting then the right-hand sides of these equations into the term 
$z_1(k)$ multiplied by $\cos(k_1l_1)\cos(k_2l_2)$, we get the right-hand 
expressions of (\ref{37}), respectively. Similarly, inserting the right-hand
sides of the relations
  \begin{equation}
k_2t_2r = 1 + {k_2t_2 \over k_1t_1}\,, ~~{k_2 \over k_1 } = k_2t_1r -{t_1 
\over t_2}\,,~~k_2 = {k_1t_1 \over t_2 (k_1t_1r -1)}\,,
\label{38b}
 \end{equation}
which follow from the same equation  $\Delta(k) =0$, 
into the term $z_2(k)$ multiplied 
by $\cos(k_1l_1)\cos(k_2l_2)$, one obtains the inverse
expressions to those in (\ref{37}), i.e., $\theta^{-1}(k)$ defined by (\ref{38}).
 Using next relations (\ref{37}) and (\ref{38}) 
 in equations (\ref{33}) and (\ref{34}),  one immediately finds
\begin{equation}
\!\!\!\!\!\!\!\!\!\!\!\!\!\!
a(k) \simeq \frac12 \!\left[\theta(k) \!+\!\theta^{-1}(k)\right]\!
 {\rm e}^{{\rm i}k(l_1 + l_2 +r)},~
b(k) \simeq \frac12\! \left[\theta(k)\! -\!\theta^{-1}(k)\right]\! 
{\rm e}^{-{\rm i}k(l_1 + l_2 +r)}. 
\label{39}
\end{equation}
One can easily see  that $\det{\bf M}=1$ and $a(k)$, $b(k)$ are well-defined 
 functions. 

In the equation $\Delta(k)=0$ [see definition (\ref{36})], all the three terms diverge in
a squeezing limit. Multiplying this equation by $r \to 0$, in the limit as
 $\gamma \to 0$, we obtain the equation 
\begin{equation}
A_1 +A_2 =A_1A_2\,,
\label{40}
\end{equation} 
where
\begin{equation}
A_j := \lim_{\gamma \to 0}(k_jt_jr) =\left\{ \begin{array}{ll} 
f_j \tan\sigma_j & \mbox{if}~~ k_jl_j  \to \sigma_j \neq 0, \\
 \eta_j & \mbox{if} ~~k_jl_j \to 0, \end{array} \right.
 \label{41}
\end{equation}
with
\begin{equation}
f_j := \lim_{\gamma \to 0}(k_jr)~~\mbox{and}~~\eta_j := 
\lim_{\gamma \to 0}(k_j^2l_jr)~~~~j=1,2.
\label{42}
\end{equation}
Here  $\sigma_j \in \R \cup \I$ ($\sigma_j$  is either 
real or imaginary and $|\sigma_j| = c_j$, the constants used for the definition 
of the ${\cal G}$-sets).   Due to definition (\ref{41}), equation (\ref{40}) 
contains  the four particular cases: (i) $\sigma_j \neq 0,~j=1,2,$  (ii)
$k_1l_1 \to 0,\, \sigma_2 \neq 0$, (iii) $\sigma_1 \neq 0,\, k_2l_2 \to 0$ and 
(iv) $k_jl_j \to 0,~j=1,2.$ The solutions to these equations define 
 the following four resonance sets, which are subsets of the ${\cal G}$-sets: 
\begin{equation}
\!\!\!\!\!\!\!\!\!\!\!\!\!\!\!\!\!\!\!
\left. \begin{array}{llll} X_{11} := \{(V_1, V_2)\in {\cal G}_{11},
  \gamma \in \Gamma_{11} \mid
f_1\tan\sigma_1 + f_2\tan\sigma_2 =f_1f_2 \tan\sigma_1\tan\sigma_2\}, \\
X_{01} := \{(V_1, V_2)\in {\cal G}_{01}, \gamma \in \Gamma_{01}  \mid \eta_1
 = (\eta_1 -1) f_2\tan\sigma_2 \},  \\
 X_{10} := \{(V_1, V_2)\in {\cal G}_{10}, \gamma \in \Gamma_{10}  \mid 
   \eta_2 =(\eta_2 -1) f_1  \tan\sigma_1 \},\\
 X_{00} := \{(V_1, V_2)\in {\cal G}_{00}, \gamma \in \Gamma_{00}  \mid 
 \eta_1 + \eta_2 =\eta_1 \eta_2\}. \end{array} \right.
\label{43}
\end{equation}
It follows from these equations that the resonance sets $ X= \{ X_{11}, X_{01}, 
X_{10}, X_{00} \}$ do not depend on $k$. According to equations (\ref{37}), 
on these sets the limit element $\theta $ is given by  
\begin{equation}
\!\!\!\!\!\!\!\!\!\!\!\!
\theta = \left\{ \begin{array}{lllll} 
(\cos\sigma_1 - f_1\sin\sigma_1) / \cos\sigma_2= \cos\sigma_1/
 (\cos\sigma_2 -f_2\sin\sigma_2) & \\ 
 ~~~~~~~~~~~~~~~~~~~= -\,f_1 \sin\sigma_1/ f_2 \sin\sigma_2\,, & X_{11}\,,\\
(1- \eta_1)/\cos\sigma_2 =1/(\cos\sigma_2 - f_2\sin\sigma_2)=
 -\,\eta_1/f_2\sin\sigma_2\,, & X_{01}\,, \\
\cos\sigma_1 - f_1 \sin\sigma_1 =\cos\sigma_1 /(1-\eta_2)= 
- f_1\sin\sigma_1/\eta_2\,, & X_{10}\,,\\
 1 -\eta_1 = 1/(1-\eta_2)= - \, \eta_1/\eta_2\,, & X_{00}\,,
 \end{array} \right. 
\label{44}
\end{equation} 
being real, which  does not depend on $k$ as well. Therefore the limit scattering data 
$a(k)$ and $b(k)$ also do not depend on $k$, so that no bound states exist
on these resonance sets. Due to asymptotic representation (\ref{39}),
the scattering data are
\begin{equation}
a = \frac12 (\theta +\theta^{-1}),~~b = \frac12 (\theta -\theta^{-1}), 
\label{45}
\end{equation}
with $\theta$ given by equations (\ref{44}). 

As follows from equations (\ref{32}), in the squeezed limit 
we have the boundary values
$\phi_1(-0) =1$, $\phi_1(+0) = a + b = \theta$ and 
$\phi'_1(-0) =- {\rm i} k$, $\phi'_1(+0) = {\rm i}k\,(b -a)=- {\rm i}k \,\theta^{-1}$. 
Similar relations take place
 for $\phi_2(x) = \phi_1^*(x)$ and thus for any function $\psi(x)$
as a linear combination of $\phi_1(x)$ and $\phi_2(x)$, the jump conditions read
\begin{equation}
\psi(+0) = \theta \,\psi(-0),~~~\psi'(+0) = \theta^{-1} \psi'(-0).
\label{z1}
\end{equation}

\subsection{The second way of the cancellation of divergences}

Here the cancellation of divergences occurs if 
 $ k_1t_1  + k_2t_2 =0 $ and  the last term in (\ref{36})
will be finite in each of the limits  $\gamma \to 0$. Then from the expression
$\Delta(k) \simeq (k_1 t_1)^2r = (k_2 t_2)^2r $ follows that the terms
$k_1t_1r^{1/2}$ and $k_2t_2r^{1/2}$ are also finite in this limit.
Therefore the terms $k_1t_1r$ and $k_2t_2r$ must disappear in any
$\gamma \to 0$ limit and therefore they can be ignored in expressions
(\ref{35}). Thus, we have the equation
\begin{equation}
\Delta(k) =-\, k_1t_1k_2t_2r  = (k_1t_1)^2r = (k_2t_2)^2r
\label{46} 
\end{equation}
and the asymptotic representation
\begin{equation}
z_1(k) \simeq 1 -(  k_1 / k_2)t_1t_2 \,,~~z_2(k) \simeq 1 -(  k_2 / k_1)t_1t_2 \,.
\label{47}
\end{equation}
Using next these relations
 in equations (\ref{33}) and (\ref{34}), one immediately finds
\begin{eqnarray}
a(k)& \simeq & \frac12 \left[\theta(k)
 +\theta^{-1}(k) + {{\rm i} \over k}\alpha(k)\right]{\rm e}^{{\rm i}k(l_1 +l_2 +r)}, 
\label{48a} \\
~b(k) &\simeq & \frac12 \left[\theta(k)
 -\theta^{-1}(k)- {{\rm i} \over k} \alpha(k)  \right]
 {\rm e}^{- {\rm i}k(l_1 +l_2 +r)}, 
\label{48}
\end{eqnarray}
where
\begin{equation}
\!\!\!\!\!\!
 \theta(k) = {\cos(k_1l_1)  \over  \cos(k_2l_2) } = 
-\,{k_1 \sin(k_1l_1) \over k_2 \sin(k_2l_2) }\,,~
\alpha(k)=k_1k_2r \sin(k_1l_1) \sin(k_2l_2) . 
\label{49}
\end{equation}

Multiplying the equation $k_1t_1 + k_2t_2 =0$ by $ r^{1/2}$, instead of 
equation (\ref{40}) we obtain the equation
\begin{equation}
B_1 +B_2 =0,
\label{50}
\end{equation}
where 
\begin{equation}
B_j := \lim_{\gamma \to 0}(k_jt_jr^{1/2}) =\left\{ \begin{array}{ll} 
g_j \tan\sigma_j & \mbox{if}~~ k_jl_j \to \sigma_j \neq 0, \\
 \beta_j & \mbox{if} ~~k_jl_j \to 0. \end{array} \right.
 \label{51}
\end{equation}
Here
\begin{equation}
g_j := \lim_{\gamma \to 0}(k_jr^{1/2})~~\mbox{and}~~\beta_j := 
\lim_{\gamma \to 0}(k_j^2l_jr^{1/2}).
\label{52}
\end{equation}
Equation (\ref{50}) together with (\ref{51}) and (\ref{52})
 defines the following four resonance sets being subsets of the ${\cal G}$-sets: 
\begin{equation}
\left. \begin{array}{llll} Y_{11} := \{ (V_1, V_2)\in {\cal G}_{11}, \gamma \in \Gamma_{11} 
 \mid g_1\tan\sigma_1 + g_2\tan\sigma_2 = 0 \}, \\
Y_{01} := \{ (V_1, V_2)\in {\cal G}_{01}, 
\gamma \in \Gamma_{01}  \mid \beta_1 + g_2\tan\sigma_2 =0 \},  \\
 Y_{10} := \{ (V_1, V_2)\in {\cal G}_{10}, \gamma \in \Gamma_{10}  \mid 
 g_1  \tan\sigma_1 +\beta_2=0 \},\\
 Y_{00} := \{ (V_1, V_2)\in {\cal G}_{00}, \gamma \in \Gamma_{00}  \mid 
 \beta_1 + \beta_2 = 0\}. \end{array} \right.
\label{53}
\end{equation}

According to (\ref{49}), on the resonance sets $Y= \{ Y_{11}, Y_{01}, Y_{10}, 
Y_{00} \}$, we have 
\begin{equation}
\theta = \left\{ \begin{array}{llll}  \cos\sigma_1/\cos\sigma_2 =
-\, g_1\sin\sigma_1/g_2 \sin\sigma_2 \,, & Y_{11}\,, \\
1/ \cos\sigma_2 = -\, \beta_1/g_2\sin\sigma_2 \,, & Y_{01}\,, \\
 \cos\sigma_1 = -\,g_1\sin\sigma_1/\beta_2\,, & Y_{10} \,, \\
 1 = -\, \beta_1/\beta_2\,, & Y_{00},
 \end{array} \right. 
 \label{54}
\end{equation}
and
\begin{equation}
\alpha =  \left\{\begin{array}{llll}
g_1 g_2 \sin\sigma_1 \sin\sigma_2\,, &~Y_{11}\,, \\
\beta_1 g_2 \sin\sigma_2\,, &~Y_{01}\,, \\
g_1 \beta_2 \sin\sigma_1\,, &~Y_{10}\,, \\
\beta_1\beta_2 \,, & ~Y_{00}\,. \end{array} \right.
\label{55}
\end{equation}
Then, according to (\ref{48a}) and (\ref{48}), we obtain
\begin{equation}
a(k) =   \frac12 \left(\theta
 +\theta^{-1} + {{\rm i} \over k}\alpha\right)\!, 
~~b(k) =  \frac12 \left(\theta -
\theta^{-1}- {{\rm i} \over k} \alpha  \right)\!, 
\label{56}
\end{equation}
where the limit elements $\theta$ and $\alpha$ are defined by equations (\ref{54})
and (\ref{55}). Similarly to (\ref{44}), the limit expressions for $\theta$ and $\alpha$
 do not depend on $k$. 
 
In the case of scattering data (\ref{56}), from  
 equations (\ref{32}) in the squeezed limit we have 
$\phi_1(-0) =1$, $\phi_1(+0) = a(k) + b(k) = \theta $ and 
$\phi'_1(-0) =- {\rm i} k$, $\phi'_1(+0) = {\rm i}k\,[b(k) -a(k)]=
- {\rm i}k \, \theta^{-1} +\alpha$. Similarly, $\phi_2(-0) =1$, 
$\phi_2(+0) = a^*(k) + b^*(k) = \theta $ and 
$\phi'_2(-0) ={\rm i} k$, $\phi'_2(+0) = {\rm i}k\,[a^*(k) -b^*(k)]=
{\rm i}k \, \theta^{-1} +\alpha$. Therefore, for  any function $\psi(x)$
from the continuum spectrum, the jump conditions read
\begin{equation}
\psi(+0) = \theta \,\psi(-0),~~~\psi'(+0) =  \theta^{-1} \psi'(-0) +
\alpha \,\phi_1(-0).
\label{z2}
\end{equation}
One can check that these equations hold true for the eigenfunctions  from
the discrete spectrum as well. Indeed, from equation $a({\rm i}\kappa )=0$
we obtain the relation $\theta + \theta^{-1} +\alpha/\kappa =0$ and,
as a result, the boundary values $\phi_1(-0) =1$, $\phi_1(+0) = b({\rm i}\kappa) 
= \theta $, $\phi'_1(-0) = \kappa$, $\phi'_1(+0) = - \kappa\, b({\rm i}\kappa)
=-\kappa \,\theta = \kappa\, \theta^{-1}  +\alpha$. Similar values take place
for $\phi_2(x)$ and $\phi_2'(x)$, so that equations (\ref{z2}) also take place
for the bound states.

\subsection{Existence of a single bound state}

Consider equations (\ref{56}).
From the equation $a({\rm i}\kappa) =0$ we find the bound state level
$\kappa$ and the value for $b(k)$ at this level:
\begin{equation}
\kappa =-\, {\alpha \over \theta +\theta^{-1}}\,,~~~b({\rm i}\kappa)=\theta.
\label{57}
\end{equation}
More explicitly, inserting expressions (\ref{54}) and (\ref{55}) into equation
(\ref{57}) and using definition (\ref{53}), the bound state level as a function 
given on the resonance $Y$-sets can be expressed through 
one of the following formulas:
\begin{eqnarray}
\kappa|_{Y_{11}} &=&   -\, {g_1 g_2 \tan\sigma_1 \tan\sigma_2 \over 
\cos^{-2}\sigma_1 + \cos^{-2}\sigma_2 } = {g_1^2 \tan^2\sigma_1 \over 
\cos^{-2}\sigma_1 + \cos^{-2}\sigma_2} \,, \nonumber \\
\kappa|_{Y_{01}} &=&  {\beta_1^2 \over 1 + \cos^{-2}\sigma_2} = 
{(\beta_1 g_2)^2 \over \beta_1^2 + 2 g_2^2}\,,\nonumber \\
\kappa|_{Y_{10}} &=&  {\beta_2^2 \over 1 + \cos^{-2}\sigma_1} = 
{( g_1 \beta_2)^2 \over 2g_1^2 + \beta_2^2 }\,, \nonumber \\
\kappa|_{Y_{00}} &=& -\, \frac12 \beta_1 \beta_2 = \frac12 \beta_1^2 
= \frac12 \beta_2^2\,. 
\label{61} 
 \end{eqnarray} 
The level $\kappa$ is an unique set function with non-trivial values only 
on the resonance $Y$-sets.

\section{Convergence of the multiple bound states of a finite-thickness
system to a  squeezed bound state}

Equations (\ref{57}) and (\ref{61})  describe  the {\it single} bound state,
which as a set function,
 is non-trivial  only on the resonance $Y$-sets. These equations
have been derived under the squeezing of a structure with finite
thickness. On the other hand, any structure  with the potential
having a well of  sufficient depth admits the existence of several
number of bound states. Therefore it would be reasonable to analyze
the behavior of all the bound states in a squeezing limit. To this end, 
we start from the equation $a({\rm i}\kappa )=0$ in which $a(k)$
is given by expression (\ref{24}), resulting in the equation $F(\kappa)=0$
with 
\begin{eqnarray}
 F(\kappa)\!\! &=& \!\!  \left[2  
+ \left( {\kappa \over k_1} -{k_1 \over \kappa} 
\right)\!t_1 + \left( {\kappa \over k_2 }
-{k_2 \over \kappa} \right)\! t_2\right]\!(1 +t_0)
\nonumber \\
\!\! \!\! &+&\!\!  \left[ \left( {\kappa^2 \over k_1k_2} + 
{ k_1k_2 \over \kappa^2 } \right)\! t_0 
-  \left( {k_1 \over k_2} +{k_2 \over k_1} \right)\right] \!t_1t_2\,,
\label{62}
\end{eqnarray}
where $t_j:= \tan(k_jl_j)$, $j=1,2$, and $t_0 := \tanh(\kappa r)$.
The function $ F(\kappa)$ is real-valued even if both $k_j$'s or one of these
 are imaginary.  As expected, no bound states exist if $V_j \ge 0$ ($j=1,2$) 
because in this case   $ F(\kappa) >0$ and  the equation 
$F(\kappa) =0$ has no solutions. The same situation takes place if
  one of $V_j$'s or both ones
are negative, but satisfy the inequalities $\kappa^2 \ge |V_j|$, $j=1,2$.
Therefore at least one of $V_j$'s has to be negative
 and then the interval of admissible
non-zero values for $\kappa$ is the interval $0 < \kappa < \max_{j=1,2}|V_j|^{1/2}$.
In other words, the solutions of the equation $F(\kappa) =0$  
have to be analyzed for the two shapes of the potential profile (\ref{21}):
(i) one of the layers  is of a barrier 
and the other one of a  well form,  and (ii) 
both the layers are of a well form.

Thus, similarly to the single-layer case \cite{d-m}, 
solving  the equation $F(\kappa) =0$ 
with respect to  $\tan(k_1l_1)$ if $V_1 \le V_2$ or 
   $\tan(k_2l_2)$ if $V_2 \le V_1$ and replacing  the variable $\kappa$
by   $\chi =k_1l_1$ (if $V_1 \le V_2$)  and $\chi =k_2l_2$ 
(if $V_2 \le V_1$) via the relations 
\begin{equation}
\kappa = \left\{ \begin{array}{ll} \sqrt{|V_1| -(\chi/l_1)^2} & 
\mbox{if}~V_1 \le V_2 \,, \\
\sqrt{|V_2| -(\chi/l_2)^2} & \mbox{if}~V_2 \le V_1\,,
\end{array} \right.
\label{63}
\end{equation}
 we arrive at the equation 
\begin{equation}
\tan\chi = y(\chi),~~~y(\chi) := { C_{0} \over  \chi C_{1} -\chi^{-1}C_{2} }\,, 
\label{64}
\end{equation}
having the same form in both the cases $V_1 \le V_2$ and $V_2 \le V_1\,.$
Here 
\begin{eqnarray}
C_0  & :=& 2 +\left( { \sqrt{\rho^2 - \chi^2} \over \zeta} 
-{\zeta \over \sqrt{\rho^2 - \chi^2} }\right) \tan\bar{\zeta}, 
\label{65} \\
C_1  &: =&  {1 \over  \sqrt{\rho^2 - \chi^2} } +\left( {1 \over \zeta} -
 {\zeta \over \rho^2- \chi^2 }\,\, t_0 \right)
 {\tan\bar{\zeta} \over 1+ t_0 }\,,
    \label{66} \\
C_2  & : =& \sqrt{\rho^2 - \chi^2} \,  - \left(  \zeta - { \rho^2 - 
\chi^2 \over \zeta}\,\,t_0 \right) {\tan\bar{\zeta} \over 1+ t_0 }\,,
 \label{67}
\end{eqnarray}
where 
\begin{equation}
\chi =\left\{ \begin{array}{ll}  k_1l_1 = l_1 \sqrt{|V_1| -  \kappa^2}
 & \mbox{if}~~V_1 \le V_2~~(V_1 <0 ) , \\
  k_2l_2 = l_2 \sqrt{|V_2| -  \kappa^2}
& \mbox{if}~~V_2 \le V_1~~(V_2 <0). \end{array} \right.
\label{68}
\end{equation}
 The other parameters in  equations (\ref{64})--(\ref{67}) are  given by 
\begin{equation}
\!\!\!\!\!\!\!\!\!\!\!\!\!\!\!
\zeta \! =\!  \left\{\begin{array}{ll} 
\zeta_1:= k_2l_1 = \! \left[\chi^2 - \rho_1^2 -V_2l_1^2  \right]^{1/2} & \mbox{if}~
V_1 \le V_2\,, \\ 
 \zeta_2:=   k_1l_2 =\! \left[\chi^2 - \rho_2^2 -V_1l_2^2\right]^{1/2} & \mbox{if}~
 V_2 \le V_1 \,, \end{array} \right.
  \bar{\zeta}\! =\! \left\{ \begin{array}{ll} \bar{\zeta}_1 := \zeta_1l_2/l_1\,, \\
 \bar{\zeta_2} :=\zeta_2l_1/l_2\,, \end{array} \right. ~
\label{69}
\end{equation}
\begin{equation}
 \rho = \left\{ \begin{array}{ll} \rho_1 := |V_1|^{1/2}l_1 & \mbox{if} ~~
V_1 \le V_2\,, \\
\rho_2 := |V_2|^{1/2}l_2 & \mbox{if} ~~V_2 \le V_1\,, \end{array} \right. 
\label{70}
\end{equation}
\begin{equation} 
 t_0  =  \left\{ \begin{array}{ll} t_{0,1} := \tanh\!\left(\sqrt{\rho_1^2 -
  \chi^2}\,r/l_1\right)
 & \mbox{if}~~V_1 \le V_2\,,\\
t_{0,2} := \tanh\!\left(\sqrt{\rho_2^2 - \chi^2}\,r/l_2\right) & 
\mbox{if}~~V_2 \le V_1\,.
 \end{array} \right.
 \label{71}
\end{equation}
The roots of equation (\ref{64}) are defined by the 
points of intersecting  the function $y(\chi)$, defined on the interval
$0 < \chi < \rho$,  with the tan-function being `standing' as $\rho$ changes. 
In such a picture, it follows  
 that the number of roots (say, $N$) is finite
and this number depends on $\rho$. 
Let us arrange the roots  in the order 
$\chi_1 >  \chi_{2} > \ldots >\chi_N$ (numbered from the right to the left). 
Correspondingly, because of the relations
\begin{equation}
\kappa_{i} =l^{-1}\sqrt{\rho^2 - \chi^2_{i}}\,\,,~~~l : =  
 \left\{ \begin{array}{ll} l_1  & \mbox{if} ~~
V_1 \le V_2\,, \\
l_2  & \mbox{if} ~~V_2 \le V_1\,, \end{array} \right. 
~ ~~i = 1, \ldots , N,
\label{74}
\end{equation}
 following from  equations (\ref{68}),
the levels $\kappa_i$'s will be arranged in the order
 $\kappa_1 < \kappa_2 < \ldots < \kappa_N.$  

\begin{figure}
\centerline{\includegraphics[width=1.\textwidth]{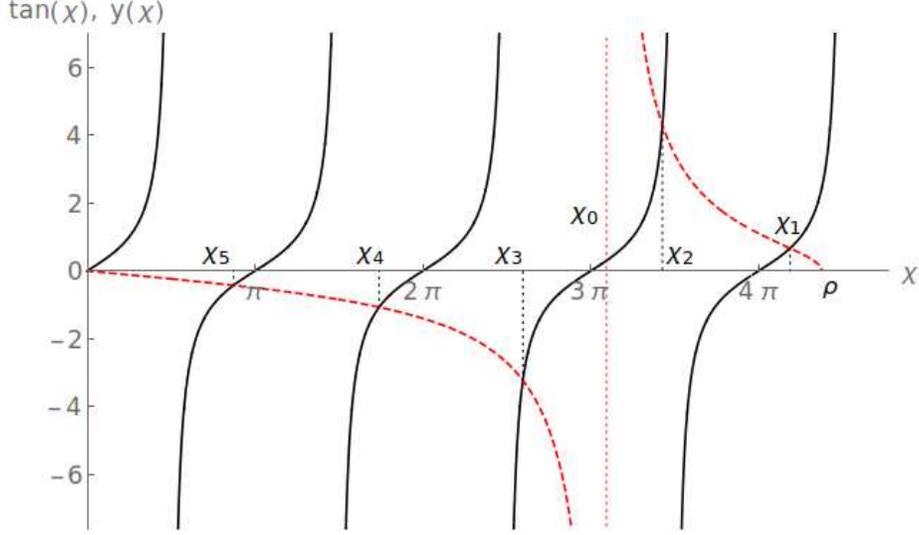}}
\caption{Graphical solution of equation (\ref{64}), where the functions 
$\tan\chi$ and $y(\chi)$ are shown by solid (black) and dashed (red)
 curves, respectively.
Five roots $\chi_1, \ldots , \chi_5$ have been obtained as a result of
 intersecting these functions. Here $\rho = 13.7$ and the function 
 $y(\chi)$ has one infinite discontinuity at point $\chi =\chi_0$. 
}
\label{fig3}
\end{figure}

Consider first the situation when one of the layers is a barrier. Then, according
to definition (\ref{69}), we have
\begin{equation}
\zeta = {\rm i}w = {\rm i}\left\{\begin{array}{ll} 
w_1:= \sqrt{  \rho_1^2 - \chi^2 + V_2l_1^2  }\,, \\ 
 w_2:=  \sqrt{  \rho_2^2 - \chi^2 + V_1l_2^2  }\,, 
 \end{array} \right.
  \tan\bar{\zeta} = {\rm i} \left\{ \begin{array}{ll} \tanh(w_1l_2/l_1), \\
 \tanh(w_2l_1/l_2),  \end{array} \right. 
\label{75}
\end{equation}
and, as a result,  the terms $C_0$, $C_1$ and $C_2$ 
in equations (\ref{65})--(\ref{67}) are positive functions of $\chi$
on the whole interval $0 < \chi < \rho$. Therefore, in the neighborhood of 
the origin $\chi =0$, the function $y(\chi)$ is negative 
 (see figure~\ref{fig3}), where $\lim_{\chi \to 0}y(\chi)=0$. 
At the other end  $\chi = \rho$, as follows from 
equations (\ref{66}) and (\ref{67}), we have $C_1 \to +\, \infty$
and $C_2 \to $ const. Then  from equation (\ref{64}) one can conclude that 
in the vicinity of the point $\chi = \rho$, the function  $y(\chi)$ 
 is positive. Since $C_0 >0$ on the whole interval $(0,\rho)$ and
  the signs of the function $y(\chi)$ in the vicinity of the ends 
$\chi=0$ and $\chi = \rho$ are opposite, the denominator of $y(\chi)$ 
has one zero. Therefore the function 
$y(\chi)$  has an infinite discontinuity
at the point $\chi_0 = \rho/\sqrt{2}$ as shown in figure~\ref{fig3}. 
On the interval $(0,\chi_0)$, the function $y(\chi)$ is negative 
and positive on the interval $(\chi_0, \rho)$. 

For the double-well (DW) form of potential (\ref{21}), some additional points of 
infinite discontinuity can appear and they will be located 
to the right of the point $\chi_0$.
  This follows from equations (\ref{69})
because at the point $\chi = l_1 \sqrt{V_2 -V_1} $ 
(if $V_1 \le V_2$) or $\chi = l_2 \sqrt{V_1 -V_2} $ (if $V_2 \le V_1$),
the parameter $\zeta$ changes from imaginary to real values. Therefore
the points of infinite discontinuity can appear on the intervals 
$l_1 \sqrt{V_2 -V_1} < \chi < \rho_1$ 
(if $V_1 \le V_2$) or $l_2 \sqrt{V_1 -V_2} < \chi < \rho_2$ (if $V_2 \le V_1$). 
In the limit as $\rho$ approaches 
the origin, the function $y(\chi)$ is negative on some interval
$0 < \chi < \chi_0$, where $\chi_0$ is the point of the first
infinite discontinuity that also approaches the origin as $\rho \to 0$.
 Therefore for sufficiently small $\rho$, only one root of equation (\ref{64})
  survives and it will be  located on the interval $(0, \pi/2)$. 
  Hence equation (\ref{64}) can admit in the $\rho \to 0$ limit only one root. 

Note that during the realization of point interactions in
 the squeezing limit,  the inequality 
$V_1(l_1) \le V_2(l_2)$ [or $V_2(l_2) \le V_1(l_1)$] must be retained,
independently on $l_1$ and $l_2$\,, because the parameter $\zeta$ in 
equations (\ref{64})--(\ref{67}) must be  either real or imaginary
during the whole squeezing procedure. Indeed, this is true because
in definition (\ref{69}) for the DW case 
we have $\chi^2 - \rho_1^2 -V_2l_1^2 =
\chi^2 - (V_1 - V_2)l_1^2$ and similarly  $\chi^2 - \rho_2^2 -V_1l_2^2 =
\chi^2 + (V_1 - V_2)l_2^2$\,, so that  the sign of 
the difference $V_1 -V_2$ is preserved during the whole squeezing process. 

Thus, for a finite double-layer structure, there exists a finite number $N$
of  solutions $\chi_1$, $\ldots , \chi_N$  to equation (\ref{64})
as illustrated by figure~\ref{fig3}.
 In the limit as $\rho \to 0$, only the
root $\chi_N$ survives. However,
in spite of the existence of this root that approaches $\rho \to 0$, 
a non-trivial limit of the level $\kappa_N$, which could follow from
the $N$th equation (\ref{74}), in general does not exist. 
Similarly, in the other limit as $\rho \to $ const. $\neq 0$, 
it follows  from equations (\ref{74}) that only the root 
$\chi_1$ can survive if it approaches $\rho$. Therefore in both these cases,
$\sqrt{\rho^2 -\chi^2} \to 0$. Taking for account this behavior, equations 
(\ref{64})--(\ref{67}) can asymptotically be replaced by 
\begin{equation}						
{\chi \tan\chi +\zeta \tan\bar{\zeta} \over  \sqrt{\rho^2 -\chi^2} } \simeq
2 - \left( {\chi \over \zeta} + {\zeta \over \chi} -
 {\chi \,\zeta \, r \over l \sqrt{\rho^2 - \chi^2} } \right) 
 \tan\chi \tan\bar{\zeta}.
 \label{76}
\end{equation}
Because of the denominator, the left-hand side of this equation diverges 
 as $\chi \to \rho$ and therefore we must  impose the condition
 \begin{equation}
\chi \tan\chi + \zeta \tan\bar{\zeta} =0.
\label{77}
\end{equation}
This is the necessary condition for the equation (\ref{76}) to be well-defined
in the limit as $\chi \to \rho$.  Coming back, according to equations 
(\ref{68})--(\ref{70}),   to the variables $k_jl_j$, $j=1,2$,
 one immediately finds that equation (\ref{77}) 
 reduces to resonance sets  (\ref{53}).
The last term in the right-hand side of equation (\ref{76})  may be finite 
as $\chi \to \rho$  if $r \to 0$  sufficiently fast. 
Setting additionally  $l \sqrt{\rho^2 - \chi_N^2} =\kappa_N$ if 
$\rho \to 0$ and $l \sqrt{\rho^2 - \chi_1^2} =\kappa_1$ if 
$\rho \to \mbox{const.} \neq 0$, we get from equation (\ref{76}) 
the same asymptotic representation $\kappa_i \simeq \Delta / (z_1 + z_2)$
with $i =1$ or $N$, where $\Delta$, $z_1$ and $z_2$ are 
defined by  equations (\ref{46}) and (\ref{47}).  
Therefore,  {\it only} on resonance sets (\ref{53}), 
 $\kappa_1$ or $\kappa_N$ converges to the bound state 
level $\kappa$ defined by equations (\ref{61}). 
 In summary, we conclude that the convergence of the multiple bound states
$\kappa_1, \ldots , \kappa_N$ to a single level $\kappa$ proceeds in
the two ways: (i) in the case if $\rho \to 0$, the lowest energy level
$\kappa_N \to \kappa$, whereas the rest of higher levels tend to zero, i.e.,
\begin{equation}
\kappa_1 \to 0,~\kappa_2 \to 0, \ldots , 
\kappa_{N-1} \to 0,~~\kappa_N \to \kappa > 0,
\label{73b}
\end{equation}
  (ii) contrary, in the case if $\rho \to \mbox{const.} \neq 0$, 
  the highest energy level $\kappa_1
\to \kappa$, while the rest lower levels escape to infinity, 
resulting in the convergence sequence
\begin{equation}
\kappa_1 \to \kappa > 0,~~~ \kappa_2, \ldots \kappa_{N} \to \infty .
\label{78}
\end{equation}
 
\section{Three-scale power-connecting parametrization of a double-layer potential}

Thus, for realizing both separated and non-separated point interactions
from a double-layer structure,
 the layer widths  $l_1, \, l_2$ and the distance between the layers $r$ 
in potential (\ref{21}) are required  to shrink  to the origin $x=0$
 along any path $\gamma = \{l_1, l_2, r \}\in \Gamma$, while $V_1$ and $V_2$
 belonging to the ${\cal G}$-sets must tend to infinity. 
 The $\Gamma$-space is interpreted as  a pencil of paths that approach the origin
obeying the limit conditions for the expressions listed in (\ref{29}) and (\ref{30}).
Only on those paths belonging to the pencil $\Gamma$, at which the resonance 
$X$- and $Y$-sets are defined by equations (\ref{43}) and (\ref{53}), the non-separated
point interactions are materialized, while beyond these sets the interactions 
are separated fulfilling boundary conditions (\ref{31}). 
Moreover, on the resonance sets, the scattering data $a(k)$ and $b(k)$
as well as bound states are shown to exist as well-defined quantities. 
From a visualization point of view, in order to demonstrate the convergence of 
the discrete spectrum and the behavior  of 
 the wave function  in a squeezed limit, it would be
convenient to parametrize potential (\ref{21}) and
 the paths $\gamma$'s via an appropriately chosen
{\it one}  squeezing parameter, which could  connect all the potential parameters
$V_1$, $V_2$, $l_1$, $l_2$ and $r$. There are various possible parametrizations
of these parameters, each of which being a subset of the ${\cal G}$-sets and
the $\Gamma$-space.
To this end, we choose  here a  power-connecting parametrization  used 
in \cite{z18aop,z10jpa,zz11jpa} for other purposes. It
couples three  positive powers $\mu $,  $\nu$ and $\tau$ via a
 dimensionless squeezing parameter $\varepsilon >0$ as follows
\begin{equation}
V_1 = \varepsilon^{-\mu}h_1\,,~V_2 = \varepsilon^{-\nu}h_2\,,~l_1= \varepsilon d_1\,,
l_2= \varepsilon^{1-\mu +\nu}d_2\,,~r =\varepsilon^\tau c.
\label{79}
\end{equation}
 Here the coefficients $h_j \in \R$, $j=1,2$, are characteristic quantities  of the system, 
so that they may be called  the layer intensities (or amplitudes). 
In the following, we
denote potential (\ref{21}) parametrized by equations (\ref{79}) as 
$V_\varepsilon(x)$. Clearly, the dependence of $V_1$ and $V_2$ on $l_1$ and $l_2$
can be expressed from (\ref{79})  explicitly using a power gymnastics. 
Then, for the limits $|V_j(l_j)|^{1/2}l_j \to c_j \ge 0$, $j =1,2$, 
to be fulfilled, the conditions on the parameters
$\mu$ and $\nu$ can be found and they will be given below.

\subsection{Existence set for the distribution $\delta'(x)$}

\begin{figure}
\centerline{\includegraphics[width=1.\textwidth]{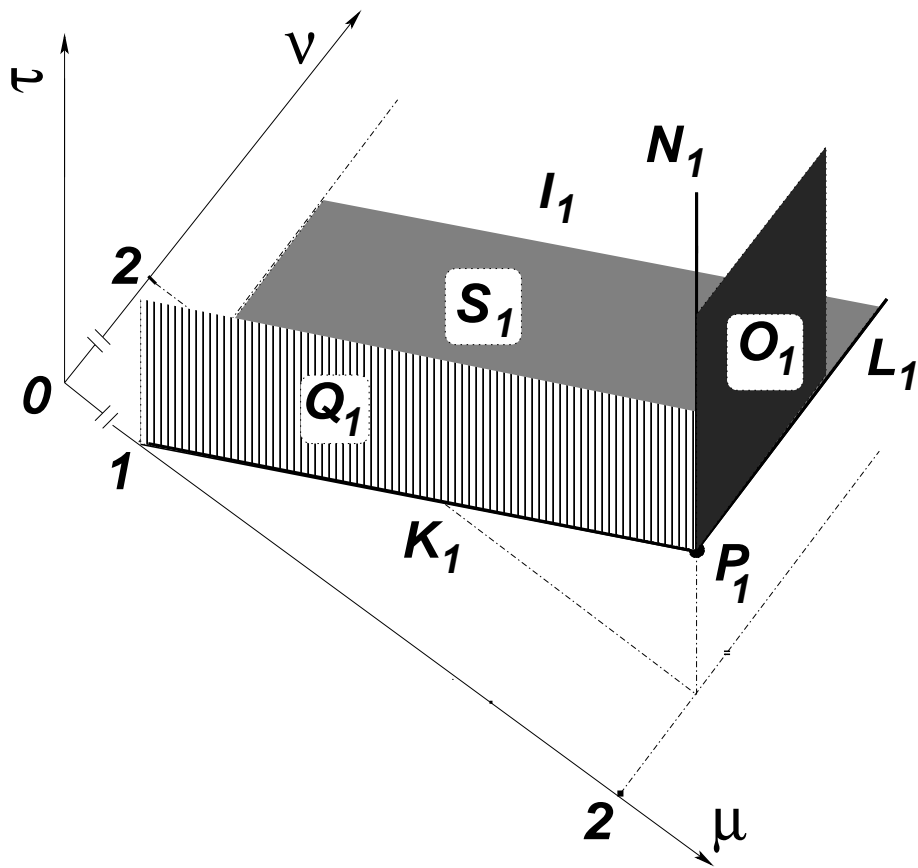}}
\caption{ The $S_{\delta'}$-surface of the existence of the distribution
$\delta'(x)$ obtained from a double-layer structure. The elements of this
surface are defined by equations (\ref{83}).
}
\label{fig4}
\end{figure}

The explicit representation (\ref{79}) allows us to define in the
$\{ \mu, \nu, \tau\}$-space the set where potential (\ref{21})
converges to $\delta'(x)$ in the sense of distributions. Thus,   
  using the fast variable $\xi = x/\varepsilon$, for any test function
   $\varphi(x) \in C_0^\infty(\R)$, we have
\begin{eqnarray}
&& \langle V_\varepsilon(x) \,| \, \varphi(x)\rangle =
\int_0^{l_1  +l_2 +r}  V_\varepsilon(x)  \varphi(x) dx \nonumber \\
&&= \varepsilon \left[ h_1 \varepsilon^{-\mu} \int_{0}^{d_1}\varphi(\varepsilon \xi)
d\xi + h_2 \varepsilon^{-\nu}\int_{d_1 + c\, \varepsilon^{\tau -1}}^{d_1 
+ d_2\, \varepsilon^{\nu -\mu}+  c\, \varepsilon^{\tau -1}}
\varphi(\varepsilon \xi)d\xi \right]. 
\label{80}
\end{eqnarray}
Expanding next $\varphi(\varepsilon \xi )= \varphi(0) + \varepsilon \xi \varphi'(0)
+ (\varepsilon \xi)^2 \varphi''(0)/2 + {\cal O}(\varepsilon^3) $, we compute
\begin{eqnarray}
\!\!\!\!\!\!\!\!\!
 \langle V_\varepsilon(x) \,| \, \varphi(x)\rangle 
& =& \varepsilon^{1-\mu} \,(h_1d_1 +h_2d_2)\,\varphi(0)
+  {\varepsilon^{2-\mu} \over 2}\left[ h_1d_1^2 + \varepsilon^{\nu -\mu} h_2d_2^2  
\right. \nonumber \\
 & +& \left. 2h_2d_2\left( d_1 +  \varepsilon^{\tau - 1} c \right) \right]
\varphi'(0) +  {\cal O}(\varepsilon^{3-\mu})+ {\cal O}(\varepsilon^{3-3\mu +2\nu})
\nonumber \\
& +& {\cal O}(\varepsilon^{\tau + 2-\mu +\nu})
 + {\cal O}(\varepsilon^{\tau +2 -2\mu +2\nu})
+ {\cal O}(\varepsilon^{2\tau +1 -\mu +\nu} ).
\label{81}
\end{eqnarray}
The first term in this expansion diverges if $\mu > 1$. However, it 
cancels out under the condition 
\begin{equation}
h_1d_1 +h_2d_2=0
\label{82}
\end{equation}
as a necessary condition for the existence of $\delta'(x)$. It can be 
fulfilled only  either on the second (WB structure) or on the fourth 
(BW structure) quadrant of the $\{h_1, h_2 \}$-plane at given 
widths $d_1$ and $d_2$.
For the analysis of the second term  in expansion (\ref{81}), in the 
$\{ \mu, \nu, \tau \}$-space  we single out
 the trihedral angle formed by vertex $ P_1$\,,  edges 
$ K_1\,, \, L_1\,, \, N_1 $ and planes $ Q_1\,, \, O_1\,, \, S_1$\,,
with interior space set $I_1$\,,
which are defined by the equations 
\begin{equation}
\begin{array}{lllllll}
  P_1 & : = & \{ \mu=\nu=2,\, \tau =1 \},\\
  K_1 & := & \{ 1<\mu <2,\, \nu =2(\mu -1),\,  \tau =\mu -1 \},\\
L_1 & := & \{ \mu =2,\, 2< \nu < \infty,\,  \tau =1 \},\\  
N_1 & := &  \{ \mu = \nu =2,\, 1< \tau < \infty \},\\
Q_1 & := & \{ 1<\mu <2,\, \nu =2(\mu -1),\,  \mu -1 < \tau < \infty \},\\
O_1 & := & \{ \mu =2,\, 2< \nu < \infty,\,   1< \tau <\infty\},\\
 S_1 & := & \{ 1<\mu <2,\, 2(\mu -1)< \nu < \infty,\,  \tau =\mu -1 \}, \\
 I_1 & := &  \{ 1<\mu <2,\, 2(\mu -1)< \nu < \infty,\,  \mu -1 < \tau < \infty \}
\end{array} 
\label{83}
\end{equation} 
and  illustrated by figure \ref{fig4}.    
In the set $I_1$\,, the second term in 
expansion (\ref{81}) vanishes as $\varepsilon \to 0$, whereas beyond 
the angle it diverges in this limit. It is remarkable that on the angle
surface $S_{\delta'} := P_1 \cup K_1 \cup L_1 \cup N_1 \cup Q_1 \cup O_1
\cup S_1$ the second term has finite limit values.
 The remainder terms in expansion (\ref{81}) tend to zero 
 in the limit as $\varepsilon \to 0 $ because all 
the powers therein are positive if they are considered 
 on the surface $S_{\delta'}$. Therefore, with taking for account 
 equation (\ref{82}), we have 
\begin{equation}
\langle V_\varepsilon(x) \,| \, \varphi(x) \rangle 
\simeq -\, {h_1d_1 \over 2}\varepsilon^{2- \mu}
\left( d_1 + \varepsilon^{\nu -\mu} d_2 + 2 \varepsilon^{\tau -1}c \right)
\varphi'(0).
\label{84}
\end{equation}
It follows from this asymptotic representation that the potential 
$V_\varepsilon(x)$ converges to $\gamma\delta'(x)$ in the sense of distributions
if condition (\ref{82}) holds true. Here,  the strength constant $\gamma$
is the set function defined by
\begin{equation}
\gamma =  {h_1 d_1 \over 2} 
\left\{ \begin{array}{lllllll}
 d_1+   d_2 +2c  \!\! & \mbox{at point}~ \, P_1\,,\\
d_2 +2c &  \mbox{on line}~ K_1\,,\\
  d_1 +2c  &  \mbox{on line}~ L_1\,,\\
d_1 + d_2  &   \mbox{on line} ~N_1\,,\\
   d_2  & \mbox{on area} ~Q_1\,,\\
d_1  & \mbox{on area} ~O_1\,, \\
 2c & \mbox{on area}~ S_1\,.
\end{array} \right. 
\label{85}
\end{equation} 
Thus, under condition (\ref{82}), the surface $S_{\delta'}$ 
separates in the $\{\mu, \nu, \tau \}$-space the volume region $I_1$ of perfect 
transmission and the region of non-existence of point interactions.

\subsection{Parametrized resonance sets and scattering data  }

\begin{figure}
\centerline{\includegraphics[width=1.\textwidth]{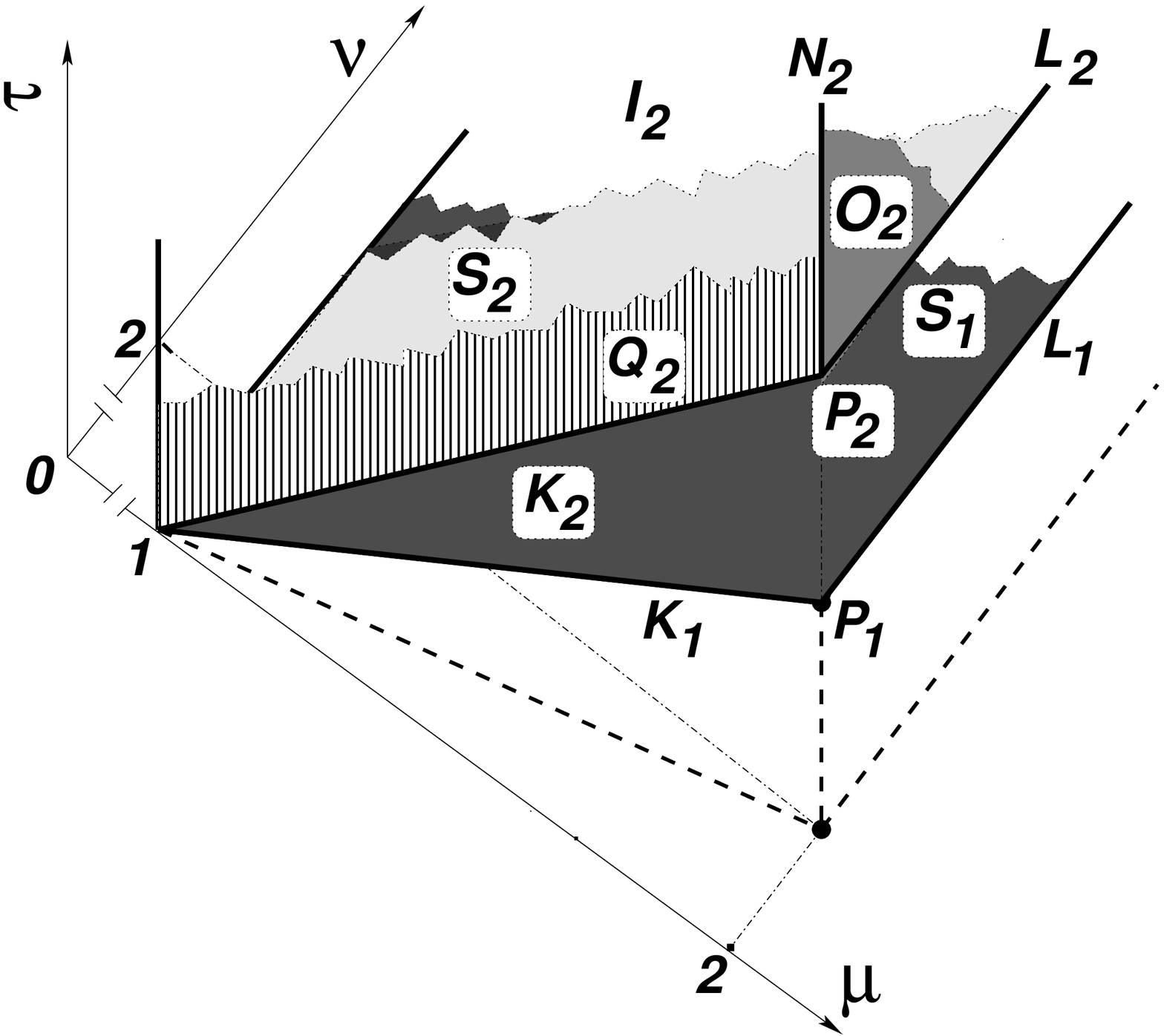}}
\caption{Sets defined by equations (\ref{83}) and (\ref{92})
 on which the point interactions of the first (with subscript `1')
and the second (with subscript `2') types are realized. The sets of the first
type are found on the plane $\{1 < \mu \le 2,~ 2(\mu -1) \le \nu < \infty,~
\tau =\mu -1 \}$, whereas the interactions of the second type in the 
space $\{1 < \mu \le 2,~ 2(\mu -1) \le \nu < \infty,~
2(\mu -1) \le \tau < \infty \}$. 
}
\label{fig5}
\end{figure}

The three-scale parametrization given by equations (\ref{79}) allows us to
realize the geometric diagram in the $\{ \mu, \nu, \tau\}$-space
(depicted in figure \ref{fig5}), where the sets for
 all the four cases of the $\varepsilon \to 0$ limits  of the arguments 
$k_1l_1$ and $k_2l_2$\,, i.e., (i) $k_1l_1 \to \sigma_1 \neq 0$
and  $k_2l_2 \to \sigma_2 \neq 0$, (ii) $k_1l_1 \to  0$
and  $k_2l_2 \to \sigma_2 \neq 0$, (iii) $k_1l_1 \to \sigma_1 \neq 0$
and  $k_2l_2 \to 0$, (iv) $k_1l_1 \to 0$
and  $k_2l_2 \to 0$, can be represented.  
The $\varepsilon \to 0$ limits of $k_jl_j$'s, $j=1,2$, define the four sets on 
the $\{\mu,\nu\}$-plane section, on which 
\begin{equation}
 k_1l_1   \simeq \varepsilon^{1- \mu/2}\sqrt{- h_1}\,d_1 \, ~~\mbox{and}~~\,
k_2l_2   \simeq \varepsilon^{1- \mu +\nu/2} \sqrt{- h_2}\,d_2 \, . 
\label{86}
\end{equation}
The other eight limits $f_j$,  $\eta_j$, $g_j$, $\beta_j$ ($j=1,2$) 
defined by equations (\ref{42}) and (\ref{52}) determine the power 
$\tau$ as a function of $\mu$ and $\nu$. Their asymptotic representation is 
\begin{eqnarray}
f_1 & \simeq & k_1r  \simeq  \varepsilon^{\tau - \mu/2} \sqrt{- h_1}\,c  , ~~
f_2 \simeq k_2r  \simeq \varepsilon^{\tau - \nu/2} \sqrt{- h_2}\,c  ,
\nonumber \\
\eta_j &\simeq & k_j^2l_jr \simeq -\, \varepsilon^{\tau -\mu +1} h_jd_jc 
\label{87}
\end{eqnarray}
for the resonance sets $X= \{ X_{11}, X_{01}, X_{10}, X_{00} \}$ and
\begin{eqnarray}
g_1 & \simeq &  k_1r^{1/2}  \simeq \varepsilon^{(\tau - \mu)/2} \sqrt{- h_1c}  , ~~
g_2 \simeq k_2r^{1/2}  \simeq \varepsilon^{(\tau - \nu)/2} \sqrt{- h_2c}  ,
\nonumber \\
\beta_j & \simeq & k_j^2l_j r^{1/2} \simeq -\, \varepsilon^{\tau/2 -\mu +1}
 h_jd_j c^{1/2}
\label{88}
\end{eqnarray}
 for the resonance sets $Y= \{ Y_{11}, Y_{01}, Y_{10}, Y_{00} \}$.
Hence  asymptotic representation (\ref{86})--(\ref{88}) admits either finite limits
\begin{equation}
\begin{array}{ll}
\sigma_j = \sqrt{- h_j}\,d_j,~~f_j = \sqrt{- h_j}\,c,~~\eta_j =-\, h_jd_j c
 & \mbox{for}~~X\mbox{-sets}, \\
\sigma_j = \sqrt{- h_j}\,d_j,~~ g_j = \sqrt{- h_jc}\,,~~
\beta_j= -\, h_jd_jc^{1/2} & \mbox{for}~~Y\mbox{-sets}, \end{array} 
\label{89}
\end{equation}
or zero. Therefore equations (\ref{43}) together with relations (\ref{89}) 
for $X$-sets are reduced to
 \begin{equation}
 \!\!\!\!\!\!\!\!\!
\left. \begin{array}{llll} X_{11}: \cot(\sqrt{-h_1}\,d_1)/\sqrt{- h_1} 
 + \cot(\sqrt{-h_2}\,d_2)/\sqrt{- h_2} \\
 X_{01}: - 1/h_1d_1 + \cot(\sqrt{-h_2}\,d_2)/\sqrt{- h_2} \\
X_{10} : \cot(\sqrt{-h_1}\,d_1)/\sqrt{- h_1} - 1/h_2d_2 \\
 X_{00}: -1/h_1d_1 - 1/h_2d_2 \end{array} \right\} =c ~~
 \left. \begin{array}{llll} \mbox{at}~~P_1\,, \\
 \mbox{on}~~K_1\,, \\
 \mbox{on}~~L_1\,, \\
 \mbox{on}~~S_1\,. \end{array} \right.
 \label{90}
 \end{equation}
Inserting next values (\ref{89}) into equations (\ref{44}),
we get the element $\theta$ and consequently the scattering data
$a$ and $b$ given by equations (\ref{45}) that do not depend on $k$.

Similarly, inserting values (\ref{89}) for $Y$-sets into equations 
(\ref{53}), we obtain 
the explicit representation of the resonance sets:
  \begin{equation}
 \!\!\!\!\!\!\!\!\!\!\!\!
\left. \begin{array}{llll} Y_{11}: \sqrt{- h_1}\tan(\sqrt{-h_1}\,d_1) 
 \! + \sqrt{- h_2}\tan(\sqrt{-h_2}\,d_2) \!\!\!\\
 Y_{01}: h_1d_1 - \sqrt{- h_2} \tan(\sqrt{-h_2}\,d_2)\!\!\!  \\
Y_{10} : \sqrt{- h_1} \tan(\sqrt{-h_1}\,d_1) - h_2d_2\!\!\! \\
 Y_{00}: h_1d_1 + h_2d_2 \!\!\! \end{array} \right\} =0 ~~\mbox{on}~
 \left\{ \begin{array}{llll} P_2\cup N_2\,, \\
 K_2\cup Q_2\,, \\
 L_2 \cup O_2\,, \\
 S_2 \cup I_2\,, \end{array} \!\! \right.~~
 \label{91}
 \end{equation}
 where the sets 
 \begin{equation} 
 \begin{array}{lllllll}
  P_2 & : = & \{ \mu=\nu=\tau =2 \},\\ 
 K_2 & := & \{ 1<\mu <2,\, \nu = \tau =2(\mu -1) \},\\
  L_2& := & \{ \mu =2,\, 2< \nu < \infty,\,  \tau =2 \},\\
  N_2 & := & \{ \mu = \nu =2,\, 2< \tau < \infty \},\\  
  Q_2 & := & \{ 1<\mu <2,\, \nu =2(\mu -1),\, \mu -1 < \tau < \infty \}, \\ 
  O_2 & := & \{ \mu =2,\, 2< \nu < \infty,\,   2< \tau <\infty \}, \\
S_2 & := & \{ 1<\mu <2,\, 2(\mu -1)< \nu < \infty,\, \tau =2(\mu -1) \},\\  
  I_2 & := & \{ 1 < \mu < 2, \, 2(\mu -1)< \nu < \infty,\, 2(\mu -1)< \tau < \infty \}
\end{array} 
\label{92}
\end{equation} 
form the second trihedral angle with vertex $P_2$, edges $K_2$, $L_2$,
$N_2$ and planes $Q_2$, $O_2$, $S_2$, including interior space set $I_2$
(see figure~\ref{fig5}). Inserting now values (\ref{89}) 
for the $Y$-sets into equations (\ref{54}) and (\ref{55}), 
we obtain the expressions for the elements $\theta$ and $\alpha$ which make
sense only in the second trihedral angle on the resonance $Y$-sets, being the
solutions to equations (\ref{91}). Note that 
these solutions (the resonance $Y_{11}$-, $Y_{01}$-, $Y_{10}$-surfaces 
and $Y_{00}$-plane) 
appear to be the limits of the corresponding $X$-surfaces as formally 
$c \to 0$ [compare equations (\ref{90}) and (\ref{91})].
 
According to asymptotic representation (\ref{88}) with limit values 
(\ref{89}) for the $Y$-sets, the elements $\theta$ given by the first 
formulas (\ref{54}) and $\alpha$ defined by equations (\ref{55})
are reduced to
\begin{equation}
\theta = \left\{ \begin{array}{llll} 
\cos(\sqrt{-\, h_1}\, d_1)/\cos(\sqrt{-\, h_2}\, d_2) & \\
1/\cos(\sqrt{-\, h_2}\, d_2) & \\
\cos(\sqrt{-\, h_1}\, d_1) & \\
1 & \\ \end{array} \right.~~~\mbox{on}~~~
\left\{ \begin{array}{llll} ~P_2 \cup N_2\,, & \\
~K_2 \cup Q_2\,, &\\
~L_2 \cup O_2\,, &\\
~S_2 \cup I_2\,, &\end{array} \right.
\label{93}
\end{equation}
\begin{equation}
\!\!\!\!\!\!\!\!\!
\alpha =  h_1h_2\, c \left\{ \begin{array}{llll}
(1 /\sqrt{- h_1}\, \sqrt{- h_2}\,) 
\sin(\sqrt{-\, h_1}\, d_1) \sin(\sqrt{-\, h_2}\, d_2)
& \mbox{at}~P_2\,,  \\
 ( d_1/\sqrt{-h_2} \,) \sin(\sqrt{-\, h_2}\, d_2) & \mbox{on}~K_2\,,  \\
 (d_2/\sqrt{-h_1}\,) \sin(\sqrt{-\, h_1}\, d_1) & \mbox{on}~L_2\,, \\ 
d_1d_2 & \mbox{on}~S_2 \end{array} \right.
\label{94}
\end{equation}
and $\alpha = 0$ in the space region $Q_2\cup O_2 \cup N_2 \cup I_2$\,.

Thus, equations (\ref{90}) and (\ref{91}) are the conditions at which 
the scattering data  $a(k)$ and $b(k)$ are well-defined quantities. 
At fixed $d_1$ and $d_2\,,$ the solutions of these  equations 
can be represented on the $\{h_1, h_2\}$-plane 
(more precisely, on the WB, DW and BW quadrants) in the form 
of curves, similarly to those depicted in the corresponding 
 figures of work \cite{z18aop} (see 
figure~2 for $X_{00}\,,$ and $Y_{00}$\,, figure~3 for $X_{01}\,,$
and $Y_{01}$\,, figure~4 for $X_{10}$ and $Y_{10}$\,,
figure~5 for $X_{11}$ and $Y_{11}$ therein). 
The cancellation of divergences of the 
first type results in the existence of resonance $X$-sets (\ref{90})
in the section plane $\tau = \mu -1$, whereas the cancellation of
the second type leads to the existence of resonance $Y$-sets
in the space region $\tau \ge 2(\mu -1)$. 

From the whole family of point
interactions, which are realized on the resonance sets described by
equations (\ref{90}) and (\ref{91}), one can single out the interactions
with  additional condition (\ref{82}) for the existence of the distribution
$\delta'(x)$. Therefore the whole family realized in general 
from both the BW and DW configurations of potential (\ref{21}) 
can be referred to as {\it generalized} $\delta'$-potentials, 
while the subfamily restricted by constraint (\ref{82}) 
{\it distributional} ones. The name `$\delta'$-potentials' (both generalized and
distributional) comes from the fact that the element $\theta(k)$ determined
by equations (\ref{44}) and (\ref{54}) on the $X$- and $Y$-sets 
except for the set $Y_{00}$ does not identically equal the unity \cite{bn}.
The point interaction with $\theta(k) =1$ [see the last equations in 
(\ref{54}) and (\ref{93})] may be called a {\it generalized} $\delta$-potential
because the corresponding profile of potential (\ref{21}) has no 
a $\delta(x)$ limit as $\varepsilon \to 0$.

\subsection{Bound state level $\kappa$}

Inserting expressions (\ref{93}) and (\ref{94}) into the formula for
the level $\kappa$ in (\ref{57}), on resonance sets (\ref{91}),
 the level $\kappa$ is 
\begin{eqnarray}
\kappa \vert_{P_2 \times Y_{11}} &= & -\, { h_1 c \tan^{2}(\sqrt{-\, h_1}\, d_1)
\over \cos^{-2}(\sqrt{-\, h_1}\, d_1)  + \cos^{-2}(\sqrt{-\, h_2}\, d_2)} 
\nonumber \\
 & =& -\, { h_2 c \tan^{2}(\sqrt{-\, h_2}\, d_2)
\over \cos^{-2}(\sqrt{-\, h_1}\, d_1)  + \cos^{-2}(\sqrt{-\, h_2}\, d_2)}\,,
\label{xxxxx} \nonumber \\
\kappa \vert_{K_2 \times Y_{01}} &= & { (h_1d_1)^2 c \over 
1 + \cos^{-2}(\sqrt{-\, h_2}\, d_2) }= {h_2 c \over 2 (h_1d_1)^{-2}h_2 -1 }\,, 
\label{yyyyy} \nonumber \\
\kappa \vert_{L_2 \times Y_{10}} &= & { (h_2d_2)^2 c \over 
1 + \cos^{-2}(\sqrt{-\, h_1}\, d_1) }= {h_1 c \over 2 (h_2d_2)^{-2}h_1 -1 }\,,
\label{zzzzz} \nonumber \\
\kappa \vert_{S_2 \times Y_{00}} &= & \frac12 (h_1d_1)^2 c = \frac12 (h_2d_2)^2 c\,.
\label{96}
\end{eqnarray}

\subsection{Convergence of multiple bound state levels to a squeezed single value}

Parametrization (\ref{79}) as a particular pathway of materializing 
point interactions from a double-layer system, 
 allows us to control explicitly the behavior 
of all the roots $\chi_1, \dots , \chi_N$ as the solutions of 
equation (\ref{64}) with (\ref{65})--(\ref{67}),  under the shrinking of the 
 system to a point. Accordingly, due to relations (\ref{74}),
one can observe the behavior of bound state levels  $\kappa_1, 
\ldots , \kappa_N$. In particular, one can establish which of the 
lateral levels $\kappa_1$ or $\kappa_N$  converges to a
single level $\kappa$. Contrary to the case with a single 
rectangular well, here the convergence is available {\it only} on the 
resonance $Y$-sets defined by equations (\ref{91}). This means that
the finite limit values for the bound state 
$\kappa$ can be obtained if  the powers $\mu$, $\nu$ and $\tau$
are found on the plane $\tau = 2(\mu -1)$, i.e., on the set
$P_2 \cup K_2 \cup L_2 \cup S_2$, on one side, and on the other side,
the system parameters $h_1,$ $d_1,$ $h_2,$ $d_2$ must obey 
equations (\ref{91}), while  $c >0$ may be arbitrary. 

Equation (\ref{64}) has been derived for the variable $\chi$ that
corresponds to the well (in the BW case) or to the 
deepest well (in the DW case). Therefore there
are  two cases: $V_1 <0$, $V_2 \in \R$, $V_1 \le V_2$ and 
$V_1 \in \R$, $V_2 <0$, $V_1 \ge V_2$.
The parameters $\rho$ as well as $\zeta$, $\bar{\zeta}$ and $t_0$
defined by equations 
(\ref{69})--(\ref{71}) and involved into equations (\ref{65})--(\ref{67})
are given explicitly in terms of $\mu, \nu, \tau$ and $\varepsilon$  as follows
\begin{equation}
\begin{array}{llll}
 \rho_1 &  = & \varepsilon^{1- \mu/2}|h_1|^{1/2}d_1\,,   \\
 \zeta_1 &  =  &\left[ \chi^2 -\rho_1^2 - \varepsilon^{2-\nu} h_2 d_1^2\, 
\right]^{1/2},\\
 \bar{\zeta}_1& =&  \left[ \varepsilon^{2(\nu -\mu)}(\chi^2 - \rho_1^2)
 (d_2/d_1)^2 - \varepsilon^{2(1-\mu) +\nu}h_2d_2^2 \right]^{1/2}, \\
 t_{0,1} & = & \tanh\!\left[\varepsilon^{\tau - 1} (c/ d_1)
\sqrt{\rho_1^2- \chi^2}\,\right]  \end{array}
\label{97}
\end{equation} 
 if $V_1 \le V_2$  and 
\begin{equation}
\begin{array}{llll}
\rho_2 & = & \varepsilon^{1- \mu +\nu/2}|h_2|^{1/2}d_2\,,  \\
\zeta_2 & = & \left[ \chi^2 - \rho_2^2 - \varepsilon^{2-3\mu+2\nu} h_1 d_2^2\, 
\right]^{1/2}, \\
\bar{\zeta}_2 & =& \left[ \varepsilon^{2(\mu -\nu)}(\chi^2 - \rho_2^2)(d_1/d_2)^2-
\varepsilon^{2-\mu} h_1 d_1^2 \right]^{1/2} ,\\
 t_{0,2} & = & \tanh\!\left[\varepsilon^{\tau - 1+\mu -\nu} 
 (c/d_2)\sqrt{\rho_2^2- \chi^2}\,\right]  \end{array}
\label{98}
\end{equation}  
if $V_2 \le V_1$. Note that one of the sets of equations 
(\ref{97}) or (\ref{98}) must be used 
during the squeezing procedure as $\varepsilon \to 0$, at least
beginning from some small value $\varepsilon > 0$.
Which of these sets has to be applied, depends on 
the shape of the potential $V_\varepsilon(x).$  The configurations 
(i), (ii) and (iii) listed below cover all 
the possible situations for the existence of bound states.
(i) WB profile ($h_1 <0,\, h_2 >0$): any $\mu$ and $\nu$, 
equations (\ref{97}) to be used.
(ii) BW profile ($h_1 >0,\, h_2 <0$): any $\mu$ and $\nu$, 
equations (\ref{98}) to be used.
(iii) DW profile ($h_1 <0,\, h_2 <0$): $h_1 \le h_2$ and $\mu =\nu$ 
or any $h_1, \, h_2$ and $\mu >\nu$, equations (\ref{97}) to
be used; $h_1 \ge h_2$ and $\mu =\nu$ 
or any $h_1, \, h_2$ and $\mu <\nu$, equations (\ref{98}) to
be used. 

 Equations (\ref{97}) and (\ref{98}) are used for the graphical 
illustration as the finite number of bound state levels $\kappa_i$'s
converges to a single level $\kappa$ given by analytic expressions 
(\ref{96}).  Plotting 
 $\tan\chi$ and $y(\chi)$ as functions of $\chi$ on the interval $0< \chi < \rho$, 
 expressed by equations (\ref{64})--(\ref{67})
in which the parameters $\zeta$, $\bar{\zeta}$, $\rho$ and $t_0$ 
 are given by one of equations (\ref{97}) or (\ref{98}), one can
find the roots $\chi_i$'s located at the intersection of the functions
$\tan\chi$ and $y(\chi)$ as shown in figure~\ref{fig3}. 
\begin{figure}
\centerline{\includegraphics[width=1.\textwidth]{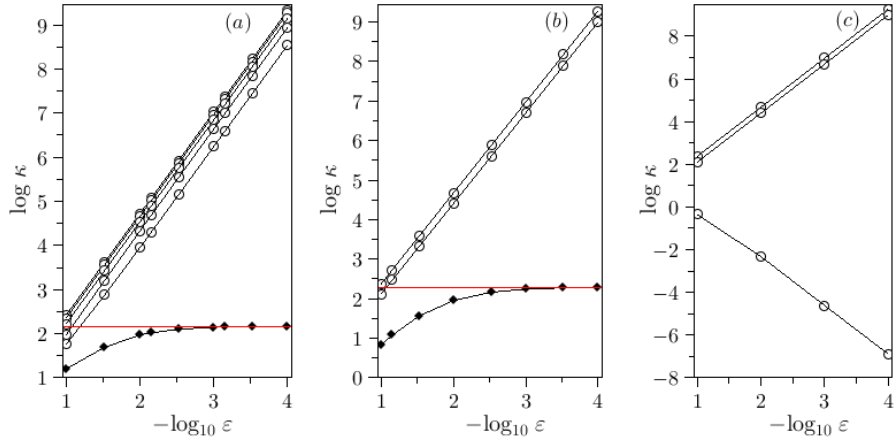}}
\caption{ Convergence of bound state levels $\kappa_i$'s in the limit as
$\varepsilon \to 0$ for the $\delta'$-potentials realized from 
double-layer structure with  parameters $h_2 = - 0.5$ eV 
and $c=20$ nm (for all cases below): 
 (a) Generalized $\delta'$-potential obtained from
 DW structure with parameters $h_1 = - 0.3$ eV,  $d_1 = 2.1$ nm and $d_2 =12.0$ nm
 obeying the first equation (\ref{91}) for $P_2$. 
 The $\{ h_1, d_1, h_2, d_2 \}$-point with these parameter values
 lies on the 1st resonance surface of the $Y_{11}$-set.
Figure~\ref{fig3} has been plotted for these values and the
five levels $\kappa_1, \ldots , \kappa_5$ as functions of $\varepsilon$
correspond to the five roots $\chi_1, \ldots , \chi_5$ at $\varepsilon =1$.
(b) Distributional $\delta'$-potential materialized from  BW structure
 with parameters $h_1 =  0.3$ eV,  $d_1 = 10.1$ nm and $d_2 = 6.1$ nm
 fulfilling both the first equation (\ref{91}) for $P_2$ 
 and constraint (\ref{82}).   The $\{ h_1, d_1, h_2, d_2 \}$-point with 
 these parameter values lies on the intersection of the 2nd resonance 
 surface of the $Y_{11}$-set  and surface (\ref{82}). For these values,
there are three roots $\chi_1, \, \chi_2, \, \chi_3$ at $\varepsilon =1$. 
The solid (red) horizontal straight lines in (a) and (b) indicate 
the values for $\kappa$ given by the formulas for 
$\kappa \vert_{P_2 \times Y_{11}}$ in (\ref{96}).
 (c) Distributional $\delta'$-potential 
 obtained from  BW profile with the same parameter values as
 in (b), but beyond the plane $\tau = 2(\mu - 1)$, lying on 
 line $N_2$ with $\tau =3$. Here, there are also three roots at 
 $\varepsilon =1$.
}
\label{fig6}
\end{figure}

The parameter values for the numerical solution of equation (\ref{64})
are chosen from the 
 $\{h_1, d_1, h_2, d_2\}$-space, which obey  equations (\ref{91}), and 
the parameter $c >0$ is supposed arbitrary. The powers  $\mu$, $\nu$ and $\tau$ 
 belong to the point $P_2$, one of the lines $K_2$
or $L_2$ and the plane $S_2 $ defined by
equations (\ref{92}) and shown in figure~\ref{fig5}.
 Having solved equation (\ref{64}) at a given $\varepsilon >0$, 
then its solutions $\chi_1, \ldots , \chi_N$
  are inserted into equations (\ref{74}) and  the   convergence of
the levels $\kappa_1, \ldots , \kappa_N$ is examined as $\varepsilon \to 0$.
Note that   the $\varepsilon \to 0$ limit values of $\kappa$ 
computed in this way  must coincide with
the analytic results given by equations (\ref{96}). As follows from
equations (\ref{97}) and (\ref{98}), in the limit as $\varepsilon \to 0$,
either $\rho_j \to 0 $ or $\rho_j \to |\sigma_j| = |h_j|^{1/2}d_j$, 
$j= 1, 2$. Below we describe the convergence of the  levels $\kappa_i$'s
on some sets of the $\{ \mu, \nu, \tau \}$-space and indicate which of
equations (\ref{97}) or (\ref{98}) has to be used in each case.
\begin{figure}
\centerline{\includegraphics[width=1.\textwidth]{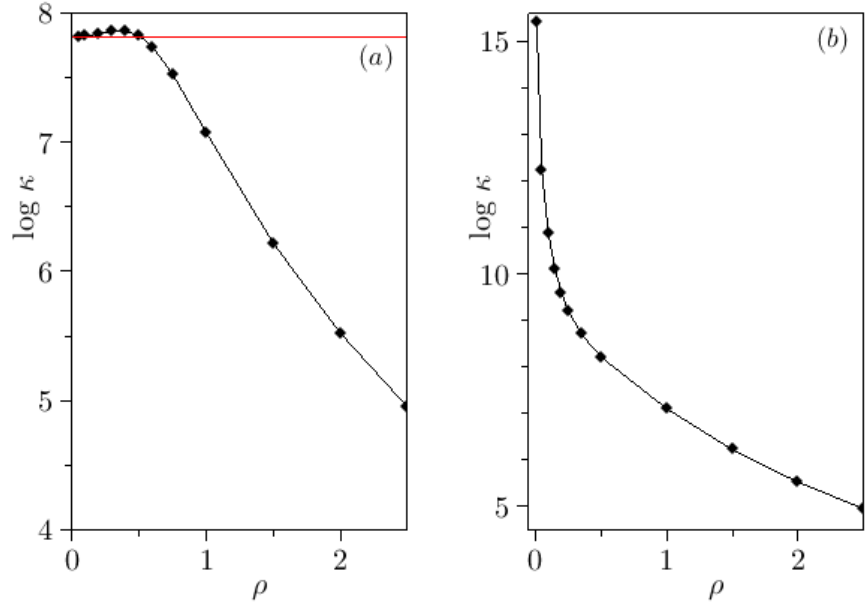}}
\caption{ Convergence of the biggest (survived) bound state level
 $\kappa_N$  as $\rho = \varepsilon^{1/4}|h_2|^{1/2}d_2 \to 0$
 for point interactions realized from BW profile with parameters 
 $h_1 = |h_2| = 0.5$ eV and $d_1 =d_2 =12$ nm: 
 (a) Generalized $\delta$-potential
  obtained for the point $\{ \mu = \nu =3/2,\, \tau = 2(\mu -1)=1\}$
  lying on plane $S_2$\,. The solid (red) horizontal straight line 
  indicates the value for $\kappa$ given by the formulas for 
  $\kappa \vert_{S_2 \times Y_{00}}$ in (\ref{96}).
   (b) Point interaction of separated type
(with full reflection) obtained for the same point on $S_2$, but 
beyond the resonance condition $h_1d_1 = |h_2|d_2 $. Here,
$d_1 = 8$ nm, $d_2 =12$ nm and $\kappa_N$ escapes to infinity. 
}
\label{fig7}
\end{figure}

{\it Point $P_2$}: At this point the convergence of $\kappa_i$, 
$i =1, \ldots , N$, is of type (\ref{78}). The generalized $\delta'$-potentials
are materialized from the WB, DW and BW configurations. Equations 
(\ref{97}) are used if $h_1 \le h_2 \in \R$ and equations (\ref{98}) if
$h_2 \le h_1 \in \R$. The distributional $\delta'$-potentials are 
realized on the WB and BW quadrants. The convergence of 
$\kappa_i$'s  to the squeezed value $\kappa$ obeying 
the first equation (\ref{96}) is illustrated in figure~\ref{fig6}: (a) for 
the generalized $\delta'$-potential obtained from a DW profile
and (b) for the distributional $\delta'$-potential realized from
a BW profile.

{\it Line $N_2$}: On this line located above the plane $\tau = 2(\mu -1)$, 
the convergence of $\kappa_i$'s is also  
of  type (\ref{78}), except for the highest energy level $\kappa_1$,
which converges to zero. In figure~\ref{fig6}(c), the convergence of this
type is plotted for the distributional $\delta'$-potential obtained 
from a BW profile.

{\it Plane $S_2$}: On this plane the family of generalized $\delta$-potentials
can be materialized if the system parameters satisfy the last equation
in (\ref{91}), which coincides with condition  (\ref{82}) 
for the existence of the distribution $\delta'(x)$. Here the convergence is
of type (\ref{73b}). The behavior of $\kappa_i$'s is shown in figure~\ref{fig7}(a)
 for the biggest level $\kappa_N$ in the particular case:
$\mu =\nu =3/2$, $h_1 = -\, h_2$, $d_1 =d_2$, where $\rho = \varepsilon^{1/4}
|h_2|^{1/2}d_2$. From  equations (\ref{74}) one finds 
the asymptotic behavior of $\kappa_N$ in the limit as $\rho \to 0$
(or $\varepsilon \to 0$) as follows
\begin{equation}
\kappa_N = \rho^{-4}\,\sqrt{\rho^2 - \chi_N^2}\, |h_2|^2d_2^3\, .
\label{d5}
\end{equation}
Using here  equations (\ref{76}) and (\ref{77}), one can arrive at
the last formula (\ref{96}). Beyond the resonance $Y_{00}$-set, 
we have that $\kappa_1 \to 0$, $\kappa_2 \to 0$, $\ldots$, $\kappa_{N-1} \to 0$,
but $\kappa_N \to \infty$ as shown in figure~\ref{fig7}(b) for the 
level $\kappa_N$.

{\it Line $K_2$}: On this line the family of generalized $\delta'$-potentials
can be  realized from the WB, DW and BW profiles on the resonance 
$Y_{01}$-set defined by the second  equation (\ref{91}). Equations (\ref{97}) 
are used with $\rho_1 \to 0$ for the WB and DW (if $h_1 \le h_2$ and $\mu =\nu$ 
or any $h_1, \, h_2$ and $\mu >\nu$) profiles. Equations (\ref{98}) 
are used with $\rho_2 \to |h_2|^{1/2}d_2$ for the
 DW (if $h_1 \ge h_2$ and $\mu =\nu$ 
or any $h_1, \, h_2$ and $\mu <\nu$) and BW profiles. In the latter case,
the distributional $\delta'$-potentials are realized if  condition 
(\ref{82}) is imposed additionally. Here
 the convergence of  type (\ref{78}) takes place 
and the behavior of $\kappa_i$'s is similar to that shown in 
figure~\ref{fig6}(b).

{\it Line $L_2$}: The situation on this line is similar to that as described
for the line $K_2$. Here the family of generalized $\delta'$-potentials
can also be  realized from the WB, DW and BW profiles, but now  on the resonance 
$Y_{10}$-set defined by the third  equation (\ref{91}). Equations (\ref{97}) 
are used with $\rho_1 \to |h_1|^{1/2}d_1$ for the WB and DW 
(if $h_1 \le h_2$ and $\mu =\nu$ 
or any $h_1, \, h_2$ and $\mu >\nu$) profiles. Equations (\ref{98}) 
are used with $\rho_2 \to 0$ for the
 DW (if $h_1 \ge h_2$ and $\mu =\nu$ 
or any $h_1, \, h_2$ and $\mu <\nu$) and BW profiles. In the  former case,
when equations (\ref{97}) have to be used,
the distributional $\delta'$-potentials are realized if  condition 
(\ref{82}) is imposed additionally. 
Here the convergence of type (\ref{78}) takes place 
and the behavior of $\kappa_i$'s is also similar to that shown in 
figure~\ref{fig6}(b).

Thus, both generalized and distributional the $\delta'$-potentials 
with non-zero bound states can be realized on the intersection of the plane 
$\tau =2(\mu - 1)$ with the surface $S_{\delta'}$, i.e., at the point $P_2$
and on the lines $K_2$ and $L_2$. The convergence of the levels $\kappa_i$'s
for the point $P_2$ is of type (\ref{78}), while on the lines $K_2$ and 
$L_2$ it can be of both types (\ref{73b}) and (\ref{78}). 
Above these sets, on the line $N_2$ and on the
planes $Q_2$ and $O_2$, the situation is quite similar, but here $\kappa_N \to 
0 $ in sequence (\ref{73b}) and $\kappa_1 \to 0$ in sequence (\ref{78}).

\subsection{Convergence of wave functions in a squeezed limit}

Parametrization (\ref{79}) of potential (\ref{21}) allows us 
to illustrate a pointwise convergence of solutions to equation (\ref{1})
in the limit as $\varepsilon \to 0$,
both for positive ($k^2 >0$) and negative ($k^2 = -\, \kappa^2$) energies.
For a non-separated point interaction to be realized, the system parameters
$h_1, h_2, d_1, d_2$ must belong to one of the resonance $X$- or $Y$-sets
described by equations (\ref{90}) and (\ref{91}). It is of interest to 
plot the wave functions on one of the $Y$-sets, because on these sets
bound states are available. Therefore, as an appropriate example, we 
choose here a BW structure with the parameter values corresponding to 
the $Y_{11}$-set and determined by the first equation in (\ref{91}).

\begin{figure}
\centerline{\includegraphics[width=1.\textwidth]{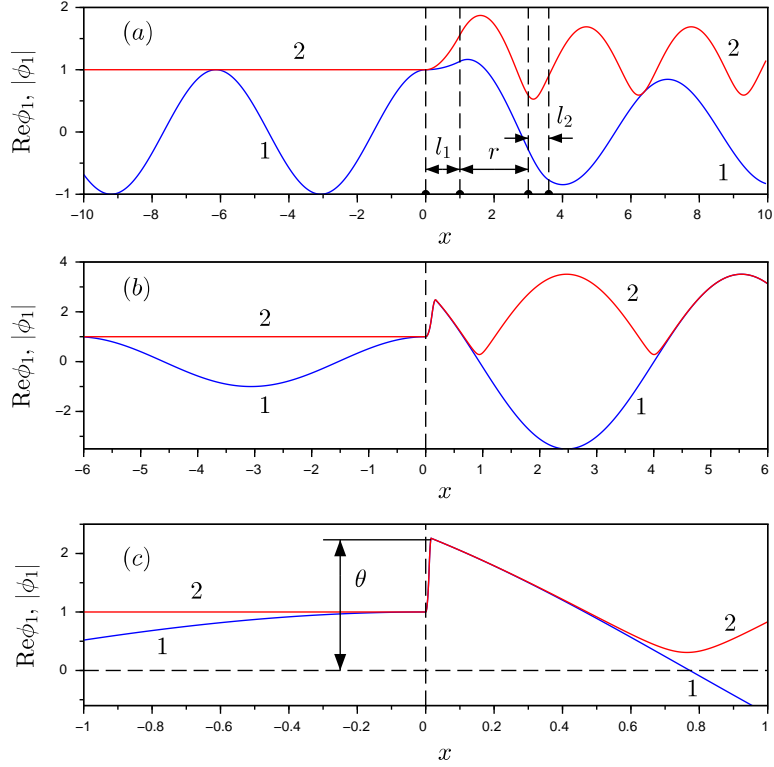}}
\caption{ Profiles    Re$\phi_1(x)$ (blue curves, 1) and 
$|\phi_1(x)|$ (red curves, 2) 
for (a) $\varepsilon =1$, (b) $\varepsilon =0.1$,
(c) $\varepsilon =0.01$. The system parameter values are $h_1 = 0.5$ eV,
$h_2 = - 0.5$ eV, $d_1 = 1.0$ nm, $d_2 = 0.6$ nm and $c = 2.0$ nm.
The energy of an incident particle is $k^2 = 0.4$ eV.
In the squeezing limit [panel (c)], the jump  of function $\phi_1(x)$
at $x=0$ reaches the value $\theta -1$ with $\theta = 2.23$.
}
\label{fig8}
\end{figure}

\begin{figure}
\centerline{\includegraphics[width=1.\textwidth]{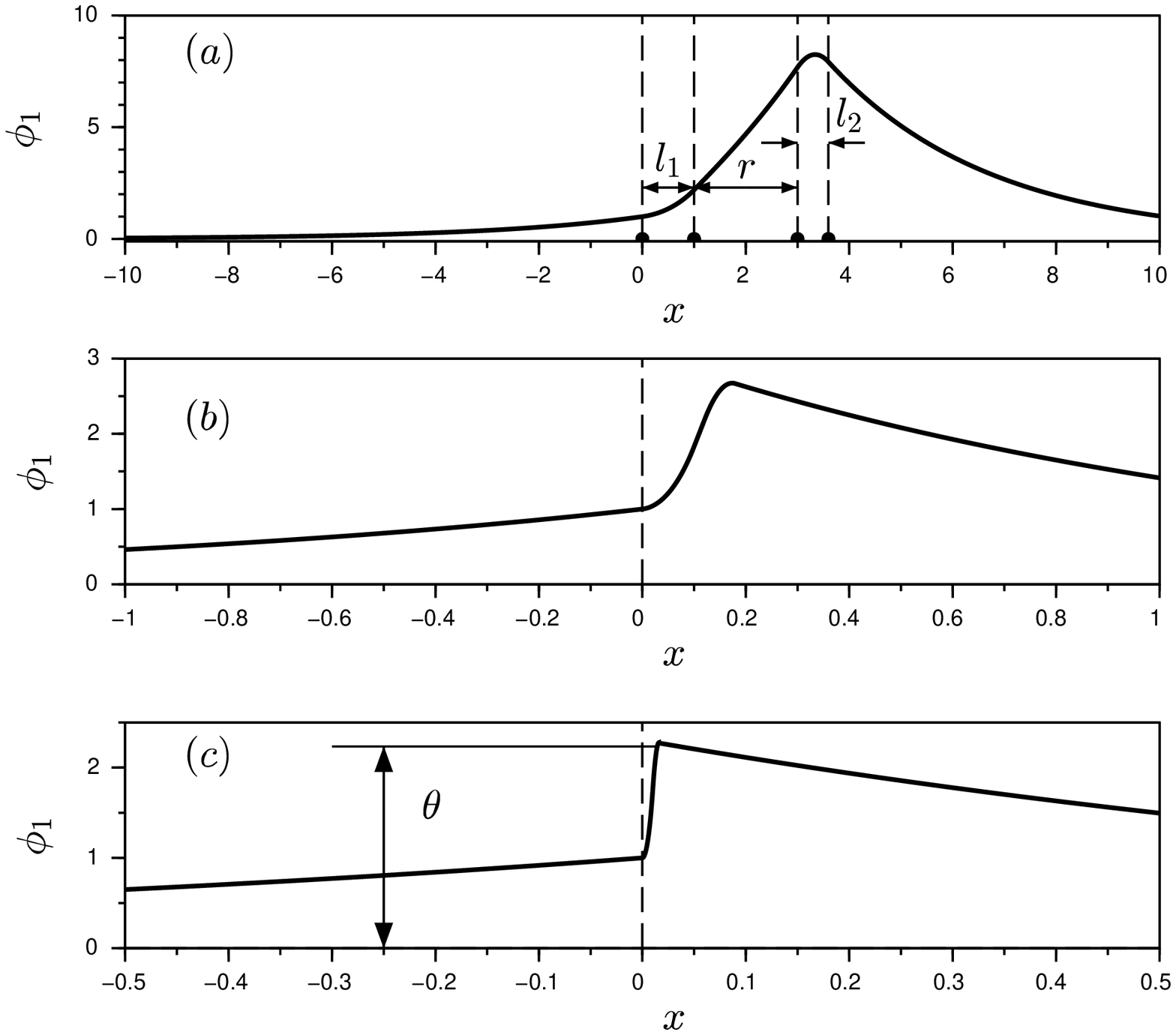}}
\caption{ Eigenfunction   $\phi_1(x)$  plotted for 
(a) $\varepsilon =1$, (b) $\varepsilon =0.1$,
(c) $\varepsilon =0.01$. The system parameter values and $\theta$ are the same as in
figure~\ref{fig8}. Here the bound state levels are $\kappa = 0.320$ 
($\varepsilon =1$), $\kappa = 0.773$ ($\varepsilon =0.1$), $\kappa = 0.864$ 
($\varepsilon =0.01$). The jump at $x=0$ illustrated by panel (c) is the same
as in figure~\ref{fig8}(c). 
}
\label{fig9}
\end{figure}

Thus, on the basis of  formula (\ref{32}), the function $\phi_1(x)$
describing an incident plane wave (with a given $k$) from the right is plotted
in figure~\ref{fig8} for three situations of shrinking a BW structure.
The parameter values $h_1, h_2, d_1, d_2$ satisfy the first equation in (\ref{91}).
For these values and the same three scales of squeezing the BW structure,
 the function $\phi_1(x)$, which describes a bound state,
is plotted in figure~\ref{fig9} using the same equations (\ref{32}) 
with $k= {\rm i}\kappa$. The panels (c) in both figures clearly illustrate
the appearance of a jump at $x=0$ that agrees with the first equation in (\ref{z2}),
where $\theta $ is computed from  the first equation in (\ref{93}). 

\bigskip

\section{Concluding remarks}

The  procedure of squeezing a double-layer structure developed in this
article is based on the simultaneous shrinking of the system parameters 
$l_1,\,\l_2$ and $r$ to zero, which can be arranged in different ways.
In this regard, the two families of the strength functions 
$V_1(l_1)$ and $V_2(l_2)$ have been defined by limit equations 
(\ref{41})--(\ref{42}) and  (\ref{51})--(\ref{52}). These equations 
 are expressed in terms of the limit characteristics 
 $\sigma_j = \lim_{l_j \to 0} (k_jl_j)$, $j=1,2$. The other eight
 limit characteristics involve the dependence on distance $r$:
$f_j , \, \eta_j \sim r^{-1}$ and  $g_j , \, \beta_j \sim r^{-1/2}$.

The squeezed limit of the scattering functions $a(k)$ and $b(k)$ 
is proven to be well-defined only if some constraints on the limit 
characteristics are imposed. These constraints are referred to as
resonance sets, resulting from the two ways of the cancellation of
divergences in the singular function $\Delta(k)$ given by formula 
(\ref{36}). The first way is to put $\Delta(k)=0$, leading to
the derivation of the resonance $X$-sets defined by equations
(\ref{43}). The second way requires that the squeezed limit of the 
function $\Delta(k)$ is a non-zero constant.  In this way, the
resonance $Y$-sets are defined by equations (\ref{53}), on which
the point interactions with non-trivial bound states can be realized. 

As a particular example of the whole variety of the shrinking ways,
 we have chosen the three-scale 
parametrization [see equations (\ref{79})], where all the system parameters
are connected through a dimensionless squeezing parameter $\varepsilon >0$. 
This connection, used in earlier publications \cite{z18aop,z10jpa,zz11jpa}, 
allows us to construct
the geometric representation in the three-dimensional space of 
positive powers $\mu$, $\nu$ and $\tau$. 

The three-scale power-connecting parametrization 
allows us to single out in the $\{ \mu, \nu, \tau \}$-space
the surface $S_{\delta'}$ of the existence of the distribution $\delta'(x)$
if condition (\ref{82}) is imposed. This condition means that only
barrier-well configurations are appropriate for the existence of 
$\delta'(x)$.  On the other hand, except for these configurations,
double-well ones also participate in realizing the point interactions
for which  $a(k)$ and $b(k)$ are well-defined functions.
In this article they are called  generalized $\delta'$-potentials.  

There exists an ubiquitous  opinion that the bound state energy levels 
  for the Schr\"{o}dinger equation (\ref{1}) with a regularized potential
  $V_\varepsilon(x)$ escape  to $-\, \infty$ as $V_\varepsilon(x)
  \to \beta \delta'(x)$ in the sense of distributions ($\beta$ is 
  a strength constant).  In this article, it is shown that 
in general this is not true,  except for the point interactions with
an additional $\delta$-like potential \cite{gnn,ggn,gggm,h,zz11jpa,g1,m-cg,gmmn},
 where $V(x)=\alpha\delta(x)
+\beta\delta'(x)$, $\alpha <0$, $\beta \in \R$. 
On the basis of both the 
analytic arguments and the numerical computations, we prove that for the family
of $\delta'$-regularized potentials with certain configurations,
 a single bound energy level converges to
a finite value, whereas the rest of energy levels escapes to $-\, \infty$. 
This is  true in general for  two families of point interactions,
called  in the present paper generalized $\delta$- and $\delta'$-potentials, 
that cover their distributional analogues. The convergence of the multiple 
bound states under shrinking the finite-thickness double-layer structure
 to a point behaves according to one of sequences (\ref{73b}) or (\ref{78}).

\bigskip
{\bf  Acknowledgments}
\bigskip

One of us (AVZ) acknowledges a partial support from the
Department of Physics and Astronomy
of the National Academy of Sciences of Ukraine (project No.~0117U000240).
 YZ acknowledges a partial support by the 
National Academy of Sciences of Ukraine Grant `Functional Properties of
Materials Prospective for Nanotechnologies' (project No.~0120U100858).
Finally, we are indebted to both Referees for suggestions, resulting in the significant
improvement of the paper.

 \end{document}